\documentclass[11pt,a4paper]{article}

\usepackage[round,colon]{natbib}
\usepackage{amsfonts,amssymb,amsmath,graphicx,epsfig,amscd,ulem}
\usepackage{graphicx}
\usepackage{array}

\def\bi{\begin{itemize}}

\def\ei{\end{itemize}}
\def\be{\begin{equation}}
\def\ee{\end{equation}}
\def\bea{\begin{eqnarray}}
\def\eea{\end{eqnarray}}
\def\th{\theta}

\def\dij{d_{ij}}
\def\drt{d_{r\theta}}
\def\dtt{d_{\theta\theta}}
\def\drr{d_{rr}}
\def\tij{\tau_{ij}}

\def\trt{\tau_{r\theta}}
\def\ttt{\tau_{\theta\theta}}

\def\gdot{\dot\gamma}
\def\bart{\bar\tau}

\def\barvt{{\bar V}_\theta}

\def\Rieq{R_{i,{\text eq}}}

\DeclareTextSymbol{\degre}{OT1}{23}

\begin{document}

\author{Guillaume Ovarlez\footnote{corresponding author: guillaume.ovarlez@lcpc.fr}, Fabien Mahaut, Fran\c cois Bertrand, Xavier Chateau\\
Universit\'e Paris Est - Laboratoire Navier (UMR 8205 ENPC-LCPC-CNRS)\\
2, all\'ee Kepler, 77420 Champs-sur-Marne, France}

\title{Flows and heterogeneities with a vane tool: MRI measurements}
\maketitle

\renewcommand{\abstractname}{Synopsis}
\begin{abstract}
We study the local flow properties of various materials in a
vane-in-cup geometry. We use MRI techniques to measure velocities
and particle concentrations in flowing Newtonian fluid, yield
stress fluid, and in a concentrated suspension of noncolloidal
particles in a yield stress fluid. In the Newtonian fluid, we
observe that the $\th$-averaged strain rate component $\drt$
decreases as the inverse squared radius in the gap, in agreement
with a Couette analogy. This allows direct comparison (without
end-effect corrections) of the resistances to shear in vane and
Couette geometries. Here, the mean shear stress in the vane-in-cup
geometry is slightly lower than in a Couette cell of same
dimensions, and a little higher than when the vane is embedded in
an infinite medium. We also observe that the flow enters deeply
the region between the blades, leading to significant extensional
flow. In the yield stress fluid, in contrast with the usually
accepted picture based on simulation results from the literature,
we find that the layer of material that is sheared near the blades
at low velocity is not cylindrical. There is thus a significant
extensional component of shear that should be taken into account
in the analysis. Finally and surprisingly, in the suspension, we
observe that a thin non-cylindrical slip layer made of the pure
interstitial yield stress fluid appears quickly at the interface
between the sheared material and the material that moves as a
rigid body between the blades. This feature can be attributed to
the non-symmetric trajectories of the noncolloidal particles
around the edges of the blades. This new important observation is
in sharp contradiction with the common belief that the vane tool
prevents slippage, and may preclude the use of the vane tool for
studying the flows of pasty materials with large particles.

\end{abstract}

\section{Introduction}\label{section_introduction}

Experimental investigations of the rheology of concentrated
suspensions often involve a vane-in-cup geometry (see
\citet{Barnes2001} for a review). The vane tool offers two main
advantages over other geometries. First, it allows the study of
the properties of structured materials with minimal disturbance of
the material structure during the insertion of the tool
[\citet{nguyen1983,alderman1991}]. It is thus widely used to study
the properties of gels and thixotropic materials
[\citet{alderman1991,Stokes2004}] and for in situ study of
materials as e.g. in the context of soil mechanics
[\citet{Richards1988}]. Second, it is supposed to avoid wall slip
[\citet{keentok1982,nguyen1983,saak2001}], which is a critical
feature in concentrated suspensions [\citet{Coussot2005}]; the
reason for this belief is that the material sheared in the gap of
the geometry is sheared by the (same) material that is trapped
between the blades. Consequently, it is widely used to study the
behavior of pasty materials containing large particles, such as
fresh concrete
[\citet{koehler2006,Estelle2008,Wallevik2008,Jau2010}] and
foodstuff [\citet{Stokes2004,Martinez2006}].

The constitutive law of materials can be obtained from a
rheological study with the vane-in-cup geometry provided one knows
the coefficients -- called ``geometry factors'' -- that allow the
conversion of the raw macroscopic data (torque, rotational angle
or velocity) into local data (shear stress, shear strain or shear
rate). However, in contrast with other classical geometries, even
the {\it a priori} simple linear problem (for Hookean or Newtonian
materials) is complex to solve with a vane tool. This linear
problem was studied theoretically by \citet{sherwood1991} and
\citet{Atkinson1992} in the general case of a $N$-bladed vane tool
embedded in an infinite linear medium. The analytical expression
found for the torque vs. rotational velocity is in rather good
agreement with macroscopic experimental data
[\citet{sherwood1991}]. Note however two possible shortcomings of
this theoretical approach for its use in practice: the blades are
infinitely thin and there is no external cylinder.

There is no such approach in the case of nonlinear media
(\textit{i.e.} complex fluids). A practical method used to study
the flow properties of non-linear materials, known as the Couette
analogy [\citet{bousmina1999,aitkadi2002,Estelle2008}], consists
in calibrating the geometry factors with Hookean or Newtonian
materials. One defines the equivalent inner radius $\Rieq$ of the
vane-in-cup geometry as the radius of the inner cylinder of a
Couette geometry that would have the same geometry factors for a
linear material. For any material, all macroscopic data are then
analyzed as if the material was sheared in a Couette geometry of
inner cylinder radius $\Rieq$. The nonlinearity (that affects the
flow field) is sometimes accounted for as it is in a standard
Couette geometry [\citet{Estelle2008}]. This approach may finally
provide constitutive law measurements within a good approximation
[\citet{Baravian2002}].

However, simulations and observations show that $\Rieq$ is not a
universal parameter of the vane tool independent of the properties
of the studied material. While the streamlines go into the virtual
cylinder delimited by the blades in the case of Newtonian media
[\citet{Baravian2002}], yielding an equivalent radius lower than
the vane radius [\citet{sherwood1991,Atkinson1992}], it was found
from simulations [\citet{Barnes1990,Savarmand2007}] that the
streamlines are nearly cylindrical everywhere for shear-thinning
fluids if their index $n$ is of order 0.5 or less, and thus that
$\Rieq=R_i$ in these cases. Moreover, for yield stress fluids,
simulations and photographs of the shearing zone around a
four-bladed vane rotating in Bingham fluids [\citet{keentok1985}],
simulations of Herschel-Bulkley and Casson fluids flows in a
four-bladed vane-in-cup geometry [\citet{Yan1997}], and
simulations of Bingham fluids flows in a six-bladed vane-in-cup
geometry [\cite{Savarmand2007}], all show that at yield
(\textit{i.e.} at low shear rates), the material contained in the
virtual cylinder delimited by the blades rotates as a rigid body,
and that it flows uniformly in a thin cylindrical layer near the
blades.  This is now widely accepted [\cite{Barnes2001}] and used
to perform a Couette analogy with $\Rieq=R_i$; the yield stress
$\tau_y$ is then simply extracted from torque $T$ measurements at
low velocity thanks to $\tau_y=T/(2\pi H R_i^2)$, where $H$ is the
vane tool height (neglecting end effects) [\citet{nguyen1992}].

The flow field in a vane-in-cup geometry and its consequences on
the geometry factors have thus led to many studies. However, only
theoretical calculations, macroscopic measurements and simulation
data exist in the literature: there are no experimental local
measurements of the flow properties of Newtonian and non-Newtonian
materials induced by a vane tool except the qualitative
visualization of streamlines made by \citet{Baravian2002} for
Newtonian media, and the photographs of \citet{keentok1985} for
yield stress fluids. Moreover, while the main advantage of the
vane tool is the postulated absence of wall slip, as far as we
know, this widely accepted hypothesis has been neither
investigated in depth nor criticized. In order to provide such
local data, we have performed velocity measurements during the
flows of a Newtonian medium and of a yield stress fluid in both a
coaxial cylinder geometry and a vane-in-cup geometry. We have also
performed particle concentration measurements in a concentrated
suspension of noncolloidal particles in a yield stress fluid,
which is a good model system for complex pastes such as fresh
concrete [\citet{Mahaut2008a,Mahaut2008b}]. Our main results are
that:

\renewcommand\theenumi{\roman{enumi}}
\renewcommand\labelenumi{(\theenumi)}

\begin{enumerate}
\item in the Newtonian fluid, the $\th$-averaged strain rate
component $\drt$ decreases as the inverse squared radius in the
gap, as in a Couette geometry, which allows direct determination
(without end-effect corrections) of the value of $\Rieq$: it is
here found to be lower than $R_i$, but slightly higher than for a
vane in an infinite medium; the flow enters deeply the region
between the blades, leading to a significant extensional flow;
\item in the yield stress fluid, in contrast with results from the
literature, the layer of material that is sheared near the blades
at low velocity does not have a cylindrical shape; \item in the
suspension of noncolloidal particles in a yield stress fluid, the
noncolloidal particles are quickly expelled from a thin zone near
the blades, leading to the development of a thin slip layer made
of the pure interstitial yield stress fluid, in sharp
contradiction with the common belief that the vane tool prevents
slippage.
\end{enumerate}

In Sec.~\ref{section_display}, we present the materials employed
and the experimental setup. We present the experimental results in
Sec.~\ref{section_results}: velocity profiles obtained with a
Newtonian oil and with a yield stress fluid are presented in
Sec.~\ref{section_results}\ref{section_oil} and
Sec.~\ref{section_results}\ref{section_emulsion}, while
Sec.~\ref{section_results}\ref{section_suspension} is devoted to
the case of suspensions, with a focus on the slip layer created by
a shear-induced migration phenomenon specific to the vane tool.

Throughout this paper, we use cylindrical coordinates $(r,\th,z)$.
All flows are supposed to be $z$ invariants (\textit{i.e.} there
are no flow instabilities). We define the $\th$-average $\bar
f(r)$ of a function $f(r,\th)$ as $\bar f(r)=(1/2\pi)\int^{2\pi}_0
f(r,\th)\,\text{d}\th $.

\section{Materials and methods}\label{section_display}

\subsection{Materials}
\begin{figure}[htbp] \begin{center}
\includegraphics[width=8cm]{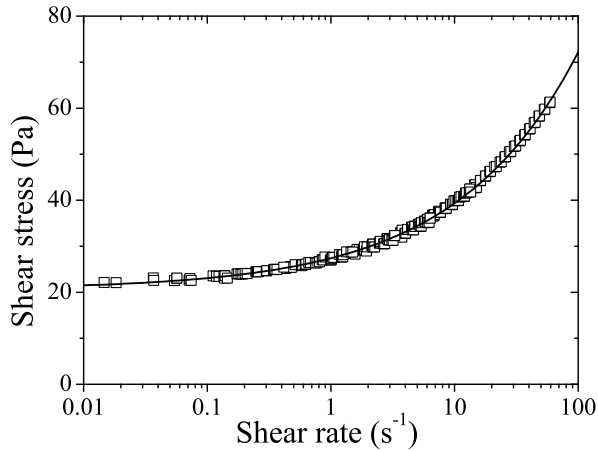}
\caption{Constitutive law of the emulsion measured locally in the
gap of a Couette cell through MRI techniques
[\citet{Ovarlez2008}]. The empty squares are local data; the solid
line is a Herschel-Bulkley fit to the data
$\tau=\tau_y+\eta_{H\!B}\gdot^{n}$ with $\tau_y=20.6$~Pa,
$\eta_{H\!B}$=6.8~Pa\,s$^{0.44}$, and
$n=0.44$.}\label{figure_emulsion_behavior}
\end{center}\end{figure}

We study three materials: a Newtonian fluid, a yield stress fluid,
and a concentrated suspension of noncolloidal particles in
this yield stress fluid.\\ The Newtonian fluid is a silicon oil of 20 mPa.s viscosity.\\
The yield stress fluid is a concentrated water in oil emulsion.
The continuous phase is dodecane oil in which Span 80 emulsifier
is dispersed at a 7\% concentration. A 100~g/l CaCl$_2$ solution
is then dispersed in the oil phase at 6000~rpm during 1 hour with
a Sliverson L4RT mixer. The droplets have a size of order 1~$\mu$m
from microscope observations. The droplet concentration is 75\%,
and the emulsion density is $\rho_f=1.01$~g\,cm$^{-3}$. The
emulsion behavior, measured through coupled rheological and MRI
techniques described in \citet{Ovarlez2008} (see
Fig.~\ref{figure_emulsion_behavior}), is well fitted to a
Herschel-Bulkley behavior $\tau=\tau_y+\eta_{H\!B}\gdot^{n}$ with
yield stress
$\tau_y=20.6$~Pa, consistency $\eta_{H\!B}$=6.8~Pa\,s$^{0.44}$, and index $n=0.44$.\\
\\
The suspension is a suspension of monodisperse polystyrene beads
(density $\rho_p=1.05$~g\,cm$^{-3}$, diameter $d=250\ \mu$m)
suspended at a 40$\%$ volume fraction in the concentrated emulsion
described above. The density matching between the particles and
the yield stress fluid is sufficient to prevent shear-induced
sedimentation of the particles in the yield stress fluid
[\citet{Ovarlez2010}]; in all experiments, we check that the
material remains homogeneous in the vertical direction by means of
MRI density measurements.

\subsection{Rheometry}
The rheometric experiments are mainly performed within a
six-bladed vane-in-cup geometry (vane radius $R_i=4.02$~cm, outer
cylinder radius $R_o=6$~cm, height $H=$11~cm). The shaft radius is
1.1~cm and the blade thickness is 6~mm. Other experiments are
performed with a wide-gap Couette geometry of slightly different
inner cylinder radius $R_i=4.15$~cm, due to the presence of
sandpaper (the other dimensions were identical). The inner
cylinder of the Couette geometry, and the outer cylinder of both
geometries are covered with sandpaper of roughness equivalent to
the size of the largest
elements of the materials studied in order to avoid wall slip.\\
In the rheometric experiments presented here, we control the
rotational velocity of the inner cylinder, with values ranging
from 0.1 to 100~rpm.

\subsection{MRI}

Proton MRI [\citet{Callaghan1991}] was chosen as a non-intrusive
technique in order to get measurements of the local velocity and
of the local bead concentration inside the sample. Experiments are
performed on a Bruker 24/80 DBX spectrometer equipped with a 0.5T
vertical superconductive magnet with 40~cm bore diameter and
operating at 21~MHz (proton frequency). We perform our experiments
with a home made NMR-compliant rheometer, equipped with the
geometries described in the previous section. This device was
already used in a number of previous rheo-nmr studies
[\citet{Raynaud2002,Rodts2004,Ovarlez2006}], and is fully
described in \citet{Raynaud2002}. The volume imaged is a (virtual)
rectangular portion of 4~cm in the axial (vertical) direction with
a width (in the tangential direction) of 1~cm and a length of 7~cm
(in the radial direction, starting from the central axis).
Velocity and concentration profiles $V_\th(r)$ and $\phi(r)$,
averaged over the vertical and tangential directions in this
volume, are obtained with a resolution of 270~$\mu$m in the radial
direction. This volume is situated at the magnet center (so as to
minimize the effects of field heterogeneities) and sufficiently
far from the bottom and the free surface of the rheometer so that
flow perturbations due to edge effects are negligible. We checked
that the velocity and concentration profiles are homogeneous along
the vertical direction in this volume, which justifies averaging
data over this direction.

Details on the sequence used to obtain velocity profiles can be
found in [\citet{Raynaud2002,Rodts2004}]. While it is possible to
get 2D or 3D maps of 2D or 3D velocity vectors
[\citet{Rodts2004}], such measurements may actually take minutes
and require complex synchronization of the MRI sequences and of
the geometry position. However, we will show in the following that
the azimuthal velocity alone provides a valuable information that
can be sufficient for most analyses; in particular it allows
computation of the $\th$-averaged strain rate component $\bar
d_{r\theta}$ and thus the $\th$-averaged shear stress
$\bart_{r\theta}$ (and the torque $T$) in the case of the
Newtonian oil. That is why we have chosen to limit ourselves to 1D
profiles of 1D velocity measurements, namely the azimuthal
velocity $V_{\theta}(r,t)$ as a function of the radius $r$ and
time $t$, for which a single measurement may take as little as
1~s; this has allowed us to perform a sufficient number of
experiments, with various materials, geometries, and rotational
velocities. Depending on the time over which this measurement is
averaged as compared to $2\pi/(N \Omega)$ where $N=6$ is the
number of blades and $\Omega$ is the rotational velocity, this
measurement may provide either a time (or $\theta$)-averaged
azimuthal velocity $\barvt(r)=(1/2\pi)\int^{2\pi}_0
V_{\theta}(r,\th) \text{d}\th $ or an instantaneous (transient)
azimuthal velocity $V_{\theta}(r,t)$. In this latter case, the
$\theta$ dependence of the azimuthal velocity
$V_{\theta}(r,\theta)$ at a given radius $r$ can then be easily
reconstructed by simply replacing the time $t$ dependence by an
angular $\theta$ dependence with $\theta=\Omega\,t$. It should
also be noted that due to incompressibility of the materials we
study, the $V_r(r,\theta)$ field can be reconstructed thanks to $
(1/r)\partial_r(rV_r)+(1/r)\partial_{\theta}(V_{\theta})=0$ with
$V_r(R_o,\theta)=0$; however, this derivation from the
experimentally measured values of $V_{\theta}(r,\theta)$ cannot be
very accurate. Finally, from $V_r(r,\theta)$ and
$V_{\theta}(r,\theta)$, we are also able to evaluate the strain
rate components $\drr(r,\th)=-\dtt(r,\th)=\partial_r V_r$ and
$\drt(r,\th)=(1/2)\big[(1/r)\partial_\theta(V_r)+r\partial_r
(V_{\theta}/r)\big]$. Note that the derivative $\partial_x f$ with
respect to coordinate $x$ of experimental data $f(x_i)$ measured
at regularly spaced positions $x_i$ was computed as: $\partial_x
f(x_i)=[f(x_{i+1})-f(x_{i-1})]/[x_{i+1}-x_{i-1}]$.

The NMR sequence used in this work to measure the local bead
concentration is a modified version of the sequence aiming at
measuring velocity profiles along one diameter in Couette geometry
[\citet{Hanlon1998,Raynaud2002}], and is described in full detail
in \citet{Ovarlez2006}. The basic idea is that during
measurements, only NMR signal originating from those hydrogen
nuclei belonging to the liquid phase of the sample (\textit{i.e.}
both the oil and water phase of the emulsions) is recorded: the
local NMR signal that is measured is thus proportional to
$1-\phi$, where $\phi$ is the local particle volume fraction. A
rather low absolute uncertainty of $\pm0.2\%$ on the concentration
measurements values was estimated in \citet{Ovarlez2006}. All
volume fraction profiles are measured at rest, after a given flow
history. This is possible because the particles do not settle in
the yield stress fluid at rest: the volume fraction profile
induced by shear is gelled by the interstitial yield stress fluid.

\section{Experimental results}\label{section_results}

In this section, we study successively the flow properties of the
Newtonian oil, the yield stress fluid, and the suspension.

\subsection{Newtonian fluids}\label{section_oil}

In this section, we study the flows observed with a Newtonian
fluid. We first present a basic theoretical analysis of the flows
in a vane-in-cup geometry as compared to flows in a standard
Couette geometry, which provides the basis for a Couette analogy.
The $\theta$-averaged azimuthal profiles $\barvt(r)$ are then
shown, and are compared to predictions of the Couette analogy. The
full velocity field $V_\theta(r,\theta)$, $V_r(r,\theta)$ is
finally presented and analyzed.

\subsubsection{Couette analogy: theoretical
analysis}\label{section_analogy}

The stress balance equation projected along the azimuthal axis is:
\bea
(1/r)\partial_r(r^2\trt)+\partial_{\theta}(\ttt)-\partial_{\theta}p=0
\label{equation_balance}\eea where $\tij$ is the deviatoric stress
tensor and $p$ the pressure.

The strain rate tensor component $\drt$ is given by: \bea
\drt(r,\th)=\frac{1}{2}\Big((1/r)\partial_\theta(V_r)+r\partial_r
(V_{\theta}/r)\Big)\label{equation_strain_rate}\eea

We recall that the constitutive law of a Newtonian fluid of
viscosity $\eta$ is: \bea\tij=2\eta
\dij\label{equation_newtonian}\eea.

In all the following analysis, we assume a no-slip boundary
condition at the walls of the inner tool and of the cup.

\subsubsection*{\it Couette geometry}

In a standard -- coaxial cylinders -- Couette geometry, due to
cylindrical symmetry, Eq.~\ref{equation_balance} becomes
$\partial_r(r^2\trt)=0$ which means that the whole shear stress
distribution $\trt(r)$ in the gap is known whatever the
constitutive law of the material is. If a torque $T(\Omega)$ is
exerted on the inner cylinder driven at a rotational velocity
$\Omega$, $\trt(r)$ is given by: \bea
\trt(r)=-\frac{T(\Omega)}{2\pi H
r^2}\label{equation_tau_Couette}\eea For a Newtonian fluid of
viscosity $\eta$, it follows that the strain rate component
$\drt(r)$ is given by: \bea \drt(r)=-\frac{T(\Omega)}{\eta 4\pi H
r^2}\eea As Eq.~\ref{equation_strain_rate} becomes
$\drt(r)=(1/2)\,r\,\partial_r (V_{\theta}/r)$ with cylindrical
symmetry, due to the boundary conditions $V_\th(R_i)=\Omega R_i$
and $V_\th(R_o)=0$, from $\int_{R_o}^{R_i}2\,\drt(r)/r\,\text{d}r
=\Omega$, one gets alternatively $\drt(r)=-\Omega R_i^2
R_o^2/[r^2(R_o^2-R_i^2)]$. This yields the following azimuthal
velocity profile: \bea V_\th(r)=\Omega
\frac{R_i^2}{r}\frac{R_o^2-r^2}{R_o^2-R_i^2}\label{equation_velocity_couette}\eea
Finally, the viscosity $\eta$ of a Newtonian fluid is obtained
from the measured torque/rotational velocity relationship
$T(\Omega)$ through
\bea\eta=\frac{T(\Omega)}{\Omega}\frac{R_o^2-R_i^2}{4\pi H R_o^2
R_i^2}\label{equation_visco_couette}\eea

These equations will be used for the comparison with the flows
observed in a vane-in-cup geometry, in particular to determine the
radius $\Rieq$ of the equivalent Couette geometry.

\subsubsection*{\it Vane-in-cup geometry}

In a vane-in-cup geometry, there is {\it a priori} no cylindrical
symmetry and all quantities {\it a priori} depend on $\th$.
However, averaging Eq.~\ref{equation_balance} over $\theta$ yields
$\partial_r(r^2\bart_{r\th})=0$. This means that what is true in a
Couette geometry, $\trt(r)=\trt(R_i)R_i^2/r^2$, is still true on
average with a vane-in-cup geometry:
$\bart_{r\th}(r)=\bart_{r\th}(R_i)R_i^2/r^2$ independently of the
material's constitutive law. Note that this derivation is true
only between $R_i$ and $R_o$; this is not true for the material
between the blades as the unknown $\tij$ distribution in the
blades contributes to the $\th$-average. The link between this
stress distribution and the torque $T(\Omega)$ exerted on the vane
tool may then seem difficult to build. However, it can be
equivalently computed on the outer cylinder as
$T=\int_0^{2\pi}\trt(R_o,\th) H R_o^2 \,\text{d}\th=2\pi H
R_o^2\bart_{r\th}(R_o)$. This means that
Eq.~\ref{equation_tau_Couette} is still valid for the
$\theta$-averaged shear stress in the vane-in-cup geometry, for
$R_i<r<R_o$: \bea \bart_{r\th}(r)=-\frac{T(\Omega)}{2\pi H
r^2}\label{equation_tau_vane}\eea

From the $\th$-averaged Eq.~\ref{equation_newtonian}, this means
that the $\th$-averaged strain rate distribution in a Newtonian
fluid, for $R_i<r<R_o$, is: \bea \bar
d_{r\th}(r)=-\frac{T(\Omega)}{\eta 4\pi H r^2}\eea From the
$\theta$-averaged Eq.~\ref{equation_strain_rate} $\bar
d_{r\theta}=(1/2)\,r\,\partial_r (\barvt/r)$, this means that the
$\th$-averaged azimuthal velocity profile of a Newtonian fluid of
viscosity $\eta$ in a vane-in-cup geometry for $R_i<r<R_o$, with a
boundary condition $\barvt(R_o)=0$ is given by: \bea
\barvt(r)=\frac{T(\Omega)}{\eta}\frac{R_o^2-r^2}{ 4\pi H R_o^2
r}\label{equation_velocity_vane}\eea

Finally, the only difference with a standard Couette flow, as
regards these $\theta$-averaged quantities, is that we do not know
the value of $\barvt(R_i)$; we only know that $
V_{\theta}(R_i,2\pi k/n)=\Omega R_i$, for $k$ integer, where $N$
is the number of blades. This means that $\bar d_{r\theta}(r)$ and
$\barvt(r)$ follow the same scaling with $r$ and $\Omega$ as in
the standard Couette geometry, but with a different prefactor.

Nevertheless, these equations provide a new insight in the Couette
analogy. The usual way of performing the Couette analogy consists
in defining the radius of the equivalent Couette geometry $\Rieq$
as the radius that allows measuring the viscosity $\eta$ of a
Newtonian fluid with the standard Couette formula. From
Eq.~\ref{equation_visco_couette}, $\eta$ should then be correctly
obtained from the torque/rotational velocity relationship
$T(\Omega)$ measured in a vane-in-cup geometry with:
\bea\eta=\frac{T(\Omega)}{\Omega}\frac{R_o^2-\Rieq^2}{ 4\pi H
R_o^2 \Rieq^2}\label{equation_visco_vane}\eea From
Eqs.~\ref{equation_visco_couette} and \ref{equation_visco_vane},
it means in particular that the torque $T_{\text{vane}}$ exerted
by the vane tool is decreased by a factor \bea
\frac{T_{\text{vane}}}{T_{\text{Couette}}}=
\frac{\Rieq^2}{R_i^2}\frac{1-R_i^2/R_o^2}{1-\Rieq^2/R_o^2}\label{equation_torque_rieq}\eea
as compared to the torque $T_{\text{Couette}}$ exerted by the
inner cylinder of a Couette geometry of same radius $R_i$ at a
same rotational velocity.

Here, from
Eqs.~\ref{equation_velocity_vane}~and~\ref{equation_velocity_couette},
we see that from the local flow perspective, there is a Couette
analogy in the sense that the $\th$-averaged azimuthal velocity
(and shear) profiles will be exactly the same as in a Couette
geometry. This defines a radius $\Rieq$ of the equivalent Couette
geometry, such that $\barvt(r)$ and $\bar d_{r\theta}(r)$ for
$R_i<r<R_o$ are given by: \bea \barvt(r)=\Omega
\frac{\Rieq^2}{r}\frac{R_o^2-r^2}{R_o^2-\Rieq^2}\label{equation_velocity_vane_eq}\\
\bar d_{r\theta}(r)=-\Omega \frac{\Rieq^2
R_o^2}{r^2(R_o^2-\Rieq^2)}\label{equation_strainrate_vane}\eea Of
course, these two definitions of $\Rieq$ are equivalent: combining
Eqs.~\ref{equation_velocity_vane} and \ref{equation_visco_vane}
yields Eq.~\ref{equation_velocity_vane_eq}.

This point of view provides an additional meaning to the Couette
analogy, namely the similarity of the average flows, and offers a
new experimental mean to determine $\Rieq$, which is more accurate
than calibration. In rheological measurements, the $T(\Omega)$
relationship has to be corrected for end effects
[\citet{sherwood1991,Martinez2007,Savarmand2007}] and the Couette
analogy has to be calibrated on a reference material of known
viscosity. Here, the $\barvt(r)$ or $\bar d_{r\th}(r)$
measurements provide the value of $\Rieq$ directly without any
correction, as only shear in the gap is involved in the analysis,
and independent of the viscosity of the material. This will be
illustrated in the following.

\subsubsection*{\it Vane in a finite cup vs. vane in an infinite medium}

The only theoretical prediction of the stress field associated
with a vane tool is that of \citet{Atkinson1992} for an infinite
$N$-bladed vane embedded in an infinite linear medium. In this
case, it is shown that the torque $ T_{\text{vane}}\left(\Omega,N,
R_{i}, R_{o}\!=\!\infty \right)$ exerted on the vane is well
approximated by \bea \frac{T_{\text{vane}}\left(\Omega,N, R_{i},
R_{o}\!=\!\infty \right)}{T_{\text{Couette}}\left(\Omega,R_{i},
R_{o}\!=\!\infty
\right)}=1-\frac{1}{N}\label{equation_prediction_atkinson} \eea
where $T_{\text{Couette}}\left(\Omega,R_{i}, R_{o}\!=\!\infty
\right)$ is the torque exerted on a cylinder of same radius $R_i$
as the vane in an infinite medium (\textit{i.e.} with
$R_{o}\!=\!\infty$). Eq.~\ref{equation_prediction_atkinson} is in
agreement with experimental results [\citet{sherwood1991}].

\citet{sherwood1991} argue that, as the stress distribution varies
as $1/r^2$ in a Couette geometry, $(R_i/R_o)^2$ should be of the
order of 1\% or less in order to nullify the influence of the
outer boundary; this is clearly the case in their experiments and
in field experiments where the vane is embedded e.g. in a soil;
this is clearly not the case in our experiments and in most
rheological experiments that make use of a vane-in-cup geometry.
However, when the cup to vane radius ratio $R_o/R_i$ is not large,
no generic theoretical expression exists in the literature.

Nevertheless, bounds of the value of the torque $T_{\text{vane}}
\left(\Omega,N, R_{i}, R_{o}\right)$ can be derived using
classical results of linear elasticity~[\cite{Salencon-2001}]. Our
starting points are the variational approaches to the solution of
the Stokes equations describing the flow of an incompressible
linear material induced by the rotation of an inner tool (of any
shape) at a rotational velocity $\Omega$ within a cup. In this
framework, it can be shown that~[\cite{Salencon-2001}]:
\begin{equation}
  \label{eq:bounds}
\int_{S_{v}}2 r \sigma^{\prime}_{\theta n}\,\text{d}S -
\frac{1}{2\Omega \eta} \int_{\omega_{\text{f}}} \tau^{\prime}_{ij}
\tau^{\prime}_{ij} \,\text{d} \omega \leq
T_{\text{vane}}\left(\Omega,N, R_{i}, R_{o}\right) \leq
 \frac{2 \eta}{\Omega}\int_{\omega_{\text{f}}} d^{\prime}_{ij} d^{\prime}_{ij} \,\text{d} \omega
\end{equation}
where $S_{v}$ denotes the inner tool-fluid interface and
$\omega_{\text{f}}$ the domain occupied by the fluid. In the first
inequality, $\sigma^{\prime}_{ij}$ is any stress tensor complying
with the stress balance equations, $\sigma^{\prime}_{\theta n}$ is
the azimuthal component of the surface forces applied by the tool
on the fluid and $\tau^{\prime}_{ij}$ is the deviatoric stress
tensor associated to $\sigma^{\prime}_{ij}$. In the second
inequality, $d^{\prime}_{ij}$ is the strain rate tensor associated
with any velocity field $\underline{V}^{\prime}$ complying with
the incompressibility constraint and the boundary conditions
prescribed on the tool-fluid and cup-fluid interfaces.

Eq.~\ref{eq:bounds} leads in particular to the expected
inequalities:
\begin{equation}
  \label{eq:expectedbounds}
 T_{\text{vane}}\left(\Omega,N, R_{i}, R_{o}\!=\!\infty \right)
\leq T_{\text{vane}}\left(\Omega,N, R_{i}, R_{o}\right) \leq
  T_{\text{Couette}}\left(\Omega, R_{i}, R_{o}\right)
\end{equation}
The lower bound is obtained by using the velocity field defined by
$\underline{V}^{\prime}=\underline{V}_{\,\Omega,N,
  R_{i}, R_{o}}$ for $r<R_o$ and $\underline{V}^{\prime} =0$ for $r
\geq R_o$, where $\underline{V}_{\,\Omega,N,
  R_{i}, R_{o}}$ is the solution for the $N$-bladed vane of radius $R_i$ in a
cup of radius $R_o$. $\underline{V}^{\prime}$ complies with the
boundary conditions for the $N$-bladed vane of radius $R_i$ in an
infinite domain problem. Then, putting this test velocity field
within the second inequality~(\ref{eq:bounds}) with $R_0=\infty$
and using
\begin{equation}
  \label{eq:clapeyron}
    T_{\text{vane}}\left(\Omega,N, R_{i}, R_{o}\right) = \frac{2
  \eta}{\Omega}\int_{\omega \left(N, R_{i}, R_{o}\right)} d_{ij} d_{ij}\,\text{d} \omega
\end{equation}
yields the lower bound of the inequality~(\ref{eq:expectedbounds})
for the quantity~$T_{\text{vane}}\left(\Omega,N, R_{i},
R_{o}\right)$. In Eq.~\ref{eq:clapeyron}, $d_{ij}$ denotes the
strain rate tensor associated with $\underline{V}_{\Omega,N,
  R_{i}, R_{o}}$ while $\omega
\left(N,
  R_{i}, R_{o}\right)$ is the domain occupied by the fluid.

The upper bound is obtained using the test velocity field defined
by $\underline{V}^{\prime} = V^{\prime}_{\theta} \left(r\right)
\underline{e}_{\theta}$ with $V^{\prime}_{\theta}$ defined by
Eq.~\ref{equation_velocity_couette} for $R_i \leq r \leq R_o$ and
by $V^{\prime}_{\theta} = \Omega r$ for $r \leq R_i$. It is easily
checked that $\underline{V}^{\prime}$ complies with the velocity
boundary conditions for any vane-in-cup geometry with vane radius
$R_i$ and cup radius $R_o$. Putting this test velocity field into
the second inequality~(\ref{eq:bounds}) then yields the upper
bound of inequality~(\ref{eq:expectedbounds}).

Finally, combining inequalities~(\ref{eq:expectedbounds}),
Eq.~\ref{equation_prediction_atkinson} and
Eq.~\ref{equation_visco_couette} yields
\begin{equation}
  \label{eq:17deGO}
  \left(1-\frac{1}{N}\right) \left(1-\frac{R_i^2}{R_o^2} \right) \leq
  \frac{T_{\text{vane}} \left(\Omega,N, R_i, R_o\right)}{T_{\text{Couette}} \left(\Omega,R_i,
    R_o\right)} \leq 1
\end{equation}

Of course, it is possible to improve the lower bound by
determining admissible test stress fields for the problem under
consideration and the inequalities~(\ref{eq:bounds}). For example,
let us consider the stress field defined between the two blades
positioned at $\theta=\pm\pi/N$ by
\begin{equation}
  \label{eq:stresstest1}
  \sigma^{\prime}_{r \theta} = - \tau \left(\frac{r}{R_i}\right)^m
\text{ ; } \sigma^{\prime}_{\theta \theta} = -(m+2) \tau \theta
\left(\frac{r}{R_i}\right)^m \text{ ; } \sigma^{\prime}_{r r} = -
\tau \theta \frac{m+2}{m+1}\left(\frac{r}{R_i}\right)^m
\end{equation}
with $m >-1$ for $r \leq R_i$ and by
\begin{equation}
  \label{eq:stresstest2}
  \sigma^{\prime}_{r \theta} = - \tau \left(\frac{r}{R_i}\right)^{-2}
\text{ ; } \sigma_{\theta \theta}=0 \text{ ; } \sigma^{\prime}_{r
r} = - \tau \theta \frac{m+2}{m+1}\left(\frac{r}{R_i}\right)^{-1}
\end{equation}
for $R_i \leq r \leq R_o$. This stress field complies with the
balance equations within the fluid domain. Let us recall that a
stress field does not need to be continuous to comply with the
balance equations (of course, this stress field is not the
solution of the problem). Putting this stress field into the first
inequality~(\ref{eq:bounds}) and using a numerical optimization
tool to choose the optimal value of the parameter $m$ yields a new
lower bound for the $N$-bladed vane in cup problem, which depends
on $N$ and $R_i/R_o$. In some cases, this test stress field
improves the lower bound of inequality~(\ref{eq:17deGO}): e.g.,
for the geometry we use in this study
($N\!=\!6,R_o/R_i\!=\!1.49$), the new lower bound is $0.57$ while
the lower bound given by inequality~(\ref{eq:17deGO}) is $0.45$.
Nevertheless, such an improvement is not obtained for all
parameter sets ($N$,$R_i/R_o$). It is thus necessary to compute
the two lower bounds for each value of ($N$,$R_i/R_o$) in order to
obtain the more accurate lower estimate of the torque. Although
application of variational approaches to the derivation of
estimates of the applied torque of a vane-in-cup problem is not
classical, it is believed that such a strategy is able to provide
useful results when no theoretical prediction of the solution is
available for particular geometries. Lower bounds of
$T_{\text{vane}}/T_{\text{Couette}}$ computed using the approach
presented above are displayed in Tab.~\ref{tab_torque_reduction}
and are compared below to our results and to data in the
literature.

\subsubsection{$\th$-averaged profiles}\label{section_mean_profiles}

\begin{figure}[htbp] \begin{center}
\includegraphics[width=7.9cm]{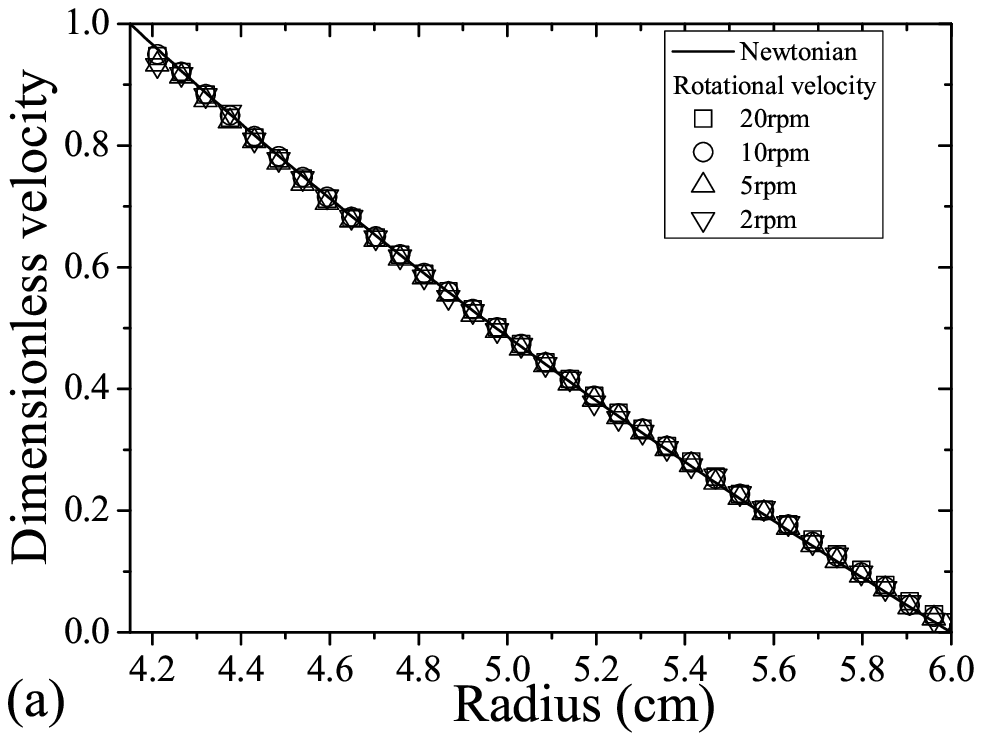}
\includegraphics[width=7.9cm]{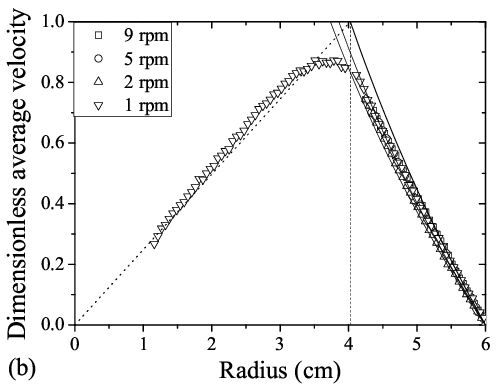}
\caption{a) Dimensionless velocity profile $V_\th(r,\Omega)/\Omega
R_i$ of a Newtonian oil in a Couette geometry ($R_i=4.15$~cm), at
various rotational velocities $\Omega$ ranging from 2 to 20~rpm;
the solid line is the theoretical profile for a Newtonian fluid.
b) Dimensionless $\th$-averaged velocity profile
$\barvt(r,\Omega)/\Omega R_i$ of a Newtonian oil in a six-bladed
vane-in-cup geometry ($R_i=4.02$~cm) for $\Omega$ ranging from 1
to 9~rpm; the vertical dashed line shows the radius of the vane;
the dotted line is the theoretical profile for a rigid body
rotation (for $r<R_i$); the solid lines are the theoretical
profiles for a Newtonian fluid in Couette geometries of radii,
from right to left: (i) $R_i=4.02$~cm, (ii) $\Rieq=3.90$~cm, and
(iii) $R_{i,th}=3.82$~cm corresponding to the \citet{Atkinson1992}
theory in an infinite medium.}\label{figure_mean_velocity_vane}
\end{center}\end{figure}

We first study the $\th$-averaged azimuthal velocity profiles
$\barvt(r)$ observed during the flows of a Newtonian oil
(Fig.~\ref{figure_mean_velocity_vane}). As shown above, these
profiles can be used to check the validity of the Couette analogy
and to determine the Couette equivalent radius $\Rieq$. The
azimuthal dependence of the velocity profiles between two adjacent
blades of the vane tool will then be considered.

In Fig.~\ref{figure_mean_velocity_vane}a we observe that the
velocity profiles in the gap of a Couette geometry are, as
expected, in perfect agreement with the theory for a Newtonian
flow (Eq.~\ref{equation_velocity_couette}). This first observation
can be seen as a validation of the measurement technique.

In the vane-in-cup geometry
(Fig.~\ref{figure_mean_velocity_vane}b), we first note that the
$\th$-averaged dimensionless azimuthal velocity profiles
$\barvt(r,\Omega)/\Omega R_i$ measured for several rotational
velocities $\Omega$ are superposed, as expected from the linear
behavior of the material. We also remark that the material between
the blades rotates as a rigid body only up to $r\simeq3.1$~cm,
indicating that the shear flow enters deeply the region between
the blades (the vane radius is 4.02~cm). The whole limit between
the sheared and the unsheared material in the ($r,\th$) plane will
be determined in
Sec.~\ref{section_results}\ref{section_oil}\ref{section_theta_profiles}
(Fig.~\ref{figure_temporal_profiles}b). We finally observe that
the theoretical velocity profile for a Newtonian fluid in a
Couette geometry of radius equal to that of the vane lies above
the data, as expected from the literature. This is also consistent
with the observation that the shear flow enters the region between
the vane blades.

\begin{figure}[htbp] \begin{center}
\includegraphics[width=9cm]{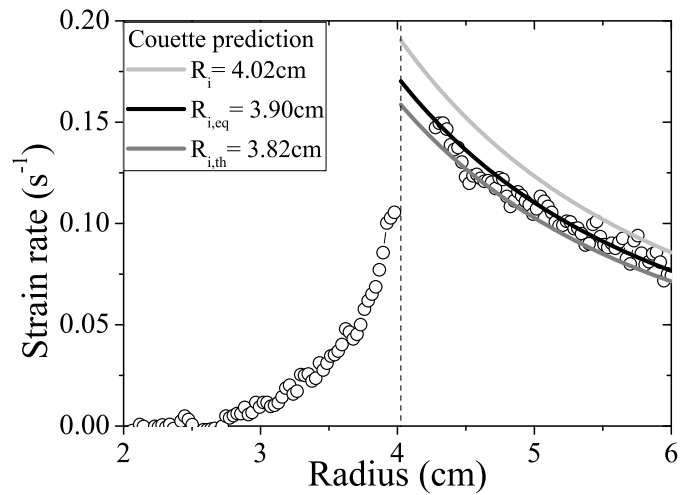}
\caption{$\th$-averaged strain rate $\bar d_{r\th}$ vs. radius $r$
for a Newtonian oil sheared at 1~rpm in a six-bladed vane-in-cup
geometry. The vertical dashed line shows the radius of the vane.
The solid lines are the theoretical strain rate profiles for a
Newtonian fluid in Couette geometries of radii: (i) $R_i=4.02$~cm
(light grey), (ii) $\Rieq=3.90$~cm (black), and (iii)
$R_{i,th}=3.82$~cm (dark grey) corresponding to the
\citet{Atkinson1992} theory in a infinite
medium.}\label{figure_mean_gradient_vane}
\end{center}\end{figure}

In order to test the Couette analogy, we have chosen to plot the
$\th$-averaged strain rate $\bar d_{r\th}$ vs. the radius $r$ in
Fig.~\ref{figure_mean_gradient_vane}. This allows us to
distinguish more clearly the difference between the experimental
and theoretical flow properties than would the velocity profiles,
because the velocity profile always tends to the same limit
($V(R_o)=0$) at the outer cylinder whereas the strain rate profile
does not. Note that velocity measurements could not be performed
close to the blades, which explains why strain rate data are
missing from 4 to 4.2~cm.

In Fig.~\ref{figure_mean_gradient_vane}, we first note that $\bar
d_{r\th}$ is zero up to $\simeq3$~cm, which corresponds to the
limit of the rigid motion of the material; $\bar d_{r\th}$ then
increases when $r$ tends towards $R_i$ as the material is more and
more sheared between the blades. In the gap of the geometry, $\bar
d_{r\th}$ decreases when $r$ increases. As expected, the
theoretical strain rate profile for a Newtonian material in a
Couette geometry of radius equal to that of the vane falls well
above the data \textit{at any radius $r$} (this was less obvious
on the velocity profiles). We then observe that the data are well
fitted to the theoretical strain rate profile
(Eq.~\ref{equation_strainrate_vane}) for a Newtonian material
flowing in an equivalent Couette geometry of inner cylinder radius
$\Rieq=3.90$~cm ($\Rieq=3.905\pm0.005$~cm was obtained from a fit
of the velocity profile to Eq.~\ref{equation_velocity_vane_eq}).
This confirms that the $\th$-averaged strain rate $\bar d_{r\th}$
decreases as the inverse squared radius in the gap, in agreement
with the Couette analogy.

From Eq.~\ref{equation_torque_rieq}, we find
$T_{\text{vane}}/T_{\text{Couette}}=0.90$ (let us recall that we
do not need to consider end effects here because we determine the
shear rate within the gap, and hence only the contribution to the
torque from the material sheared in the gap). This value can now
be compared to data from the literature. For a six-bladed vane
tool in an infinite medium, the \citet{Atkinson1992} theory would
imply a theoretical $T_{\text{vane}}/T_{\text{Couette}}=0.83$,
which is 8\% lower than what we measure, and corresponds to a
theoretical ``equivalent radius'' $R_{i,th}=3.82$~cm when the vane
is embedded in a cup of radius $R_o=6$~cm.
Figs.~\ref{figure_mean_velocity_vane}b and
\ref{figure_mean_gradient_vane} show that the flow characteristics
predicted with this value of the equivalent radius can be
distinguished from our experimental data and fall slightly below
the data (a discrepancy could be expected as $(R_i/R_o)^2$ is not
small in our experiment).

\begin{table}[htbp] \begin{center}\begin{tabular}{|c|c|c|c|c|c|c|} \hline
Study & $N$ & $R_o/R_i$ & $r_s/R_i$ & $\epsilon/R_i$ & $\frac{T_{\text{vane}}}{T_{\text{Couette}}}$ & Lower \\
&&&&&& bound\\
\hline
\citet{Atkinson1992} (th.) & 4 & $\infty$ & 0 & 0 & 0.75 & -- \\
\citet{Zhang1998} (num.) & 4 & 2 & -- & -- & 0.65 & 0.56 \\
\citet{Martinez2007} (exp.) & 4 & 1.825 & 0.6 & 0.045 & 0.67 & 0.52  \\
\citet{Zhu2010} (num.) & 4 & 1.14 & 0.14 & 0.1 & 0.73 & 0.26 \\
\citet{Barnes1990} (num.) & 4 & 1.12 & 0.35 & 0.12 & 0.61 & 0.24 \\
\hline
\citet{Atkinson1992} (th.) & 6 & $\infty$ & 0 & 0 & 0.83 & -- \\
\cite{Baravian2002} (exp.) & 6 & 2.55 & -- & -- & 0.7 & 0.7 \\
\citet{Zhang1998} (num.) & 6 & 2 & -- & -- & 0.74 & 0.63 \\
Present study (exp.) & 6 & 1.49 & 0.27 & 0.15 & 0.90 & 0.57 \\
\hline
\end{tabular}\caption{Ratio $T_{\text{vane}}/T_{\text{Couette}}$ between the torque
measured when straining a linear medium (viscous or elastic) in a
vane-in-cup geometry and that measured in a coaxial cylinder
geometry of similar dimensions, obtained in various theoretical,
numerical and experimental studies of the literature; only data
corrected for (or free from) end effects are shown. The number $N$
of blades, the cup to vane radius ratio $R_o/R_i$, the shaft
radius to vane radius ratio $r_s/R_i$, and the blade thickness to
vane radius ratio $\epsilon/R_i$, are displayed when provided in
the manuscripts. The theoretical lower bound computed using
variational approaches in
Sec.~\ref{section_results}\ref{section_oil}\ref{section_analogy}
is also
provided.}\label{tab_torque_reduction}\end{center}\end{table}

We have gathered experimental and numerical data from the
literature where the cup to vane radius ratio $R_o/R_i$ is not
large in Tab.~\ref{tab_torque_reduction}; only data corrected for
(or free from) end effects are shown. First, it should be noted
that all torque data obey the theoretical inequalities computed
using variational approaches in
Sec.~\ref{section_results}\ref{section_oil}\ref{section_analogy}.
However, no clear trends emerge from the comparison of the data.
The relative impact of the various geometrical parameters that may
affect the flow field, namely the cup to vane radius ratio
$R_o/R_i$, the shaft radius to vane radius ratio $r_s/R_i$, and
the blade thickness to vane radius ratio $\epsilon/R_i$, cannot be
determined at this stage. For example, in very similar geometries,
\citet{Zhu2010} find a torque ratio
$T_{\text{vane}}/T_{\text{Couette}}$ close to that of
\citet{Atkinson1992} whereas \citet{Barnes1990} find a much lower
torque ratio. The only noticeable difference between these two
studies (apart from the numerical method) is that $r_s/R_i$ is
higher in \citet{Barnes1990}, but our data, with a rather large
value of $r_s/R_i$, show different features. We actually note that
our study is the only one to report a torque ratio higher than in
an infinite medium; all other data report torque ratios up to 19\%
lower than expected in an infinite medium. In the general case of
a finite vane-in-cup geometry, it thus seems that numerical
investigations are still needed, and that, at this stage, a
calibration has to be performed to get the geometry factors. We
also expect that the bounds obtained using variational approaches
in
Sec.~\ref{section_results}\ref{section_oil}\ref{section_analogy}
can be improved.

\subsubsection{$\th$ dependent profiles}\label{section_theta_profiles}

To better characterize the flow field, we now study the dependence
of the velocity profiles on the angular position $\th$. We have
performed experiments in which we measure one azimuthal velocity
profile per second while the vane tool is rotated at 1~rpm,
yielding 10 profiles between two adjacent blades.

\begin{figure}[htbp] \begin{center}
\includegraphics[width=8.7cm]{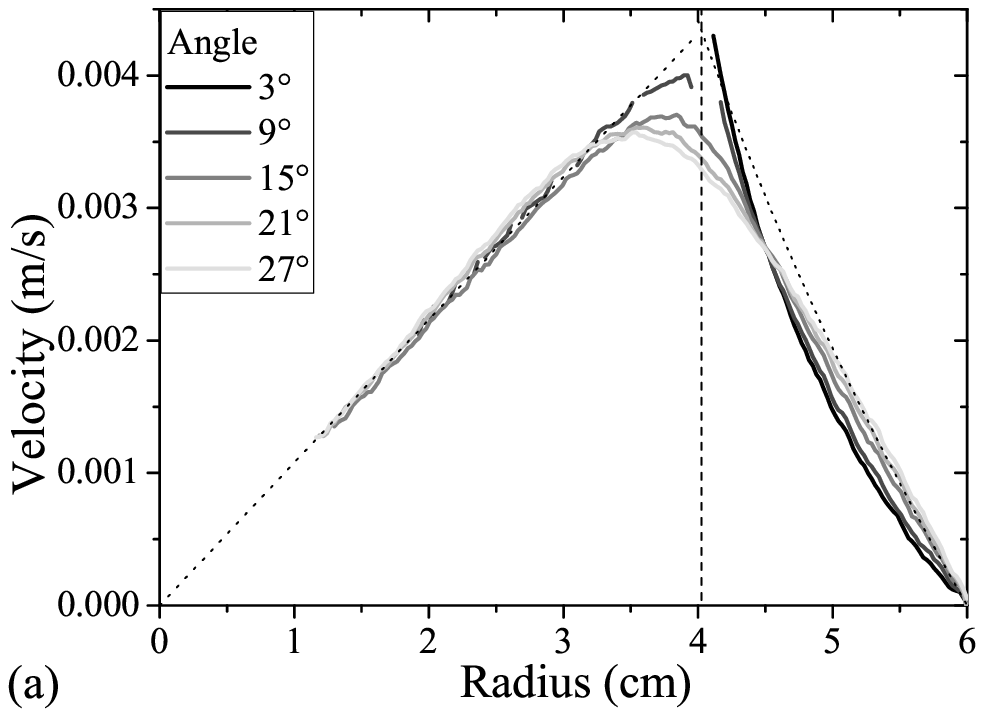}
\includegraphics[width=7.1cm]{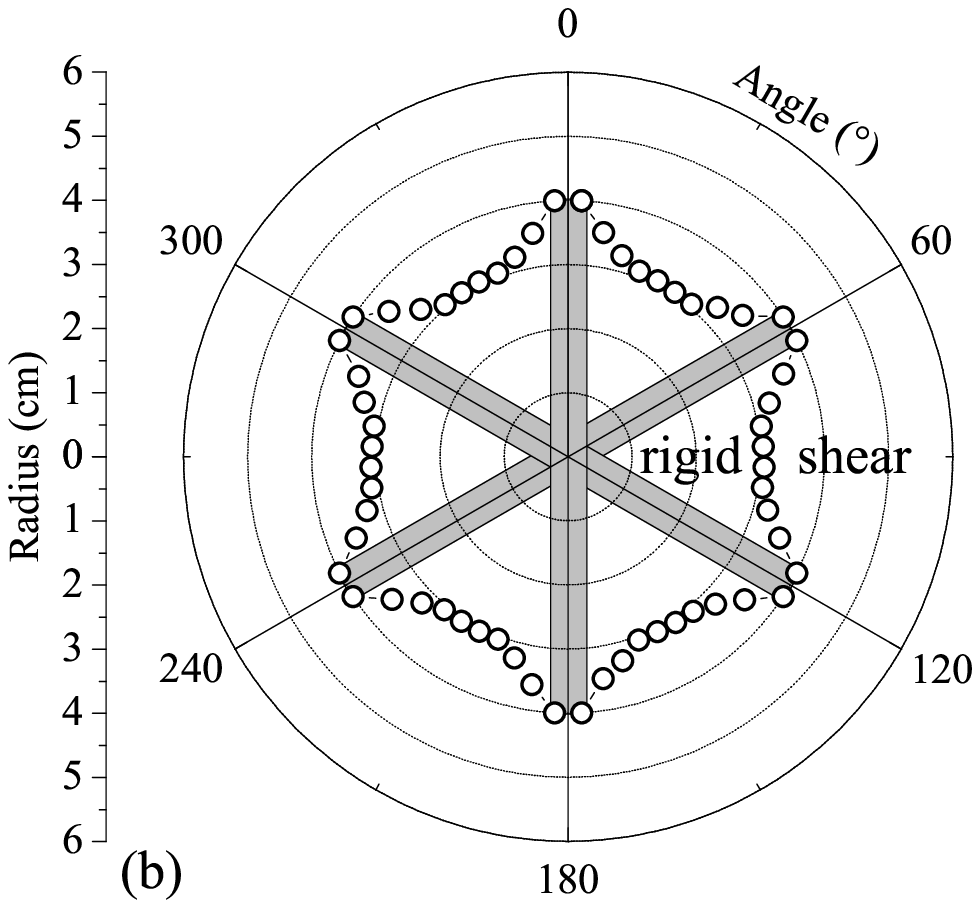}
\caption{a) Azimuthal velocity profile $V_\th(r,\th)$ of a
Newtonian oil sheared at 1~rpm in a six-bladed vane-in-cup
geometry, for various angular positions, $\th$, between one blade
($\th=0$\degre) and midway between adjacent blades
($\th=30$\degre). The vertical dashed line shows the radius of the
vane. The dotted line is the profile for a rigid body rotation
(for $r<R_i$) and the theoretical profile for a Newtonian fluid in
a Couette geometry of radius $R_i$ (for $r>R_i$). b)
Two-dimensional plot of the limit between rigid motion and shear
(empty circles) for a Newtonian material in the six-bladed
vane-in-cup geometry; the grey rectangles correspond to the
blades.}\label{figure_temporal_profiles}
\end{center}\end{figure}

In Fig.~\ref{figure_temporal_profiles}a, we plot the velocity
profiles $V_\th(r,\th)$ measured at different angles $\th$. We
first observe that the velocity profile which starts near from a
blade tip (corresponding to $\th=0$\degre by definition) is very
different from the velocity profile in a Couette geometry of same
radius: it starts with a much steeper slope, which means that the
blades tip neighborhoods are regions of important shear as already
observed by \citet{Barnes1990}. We then observe that, as expected
from the $\th$-averaged velocity profiles, the shear flow enters
more and more deeply the region between the blades as $\th$ tends
towards $30$\degre (corresponding to midway between two adjacent
blades); at this angular position, the rigid rotation stops at
$R_l\simeq3.05$~cm. From all the velocity profiles, we finally
extract a 2D map of the limit $R_l(\th)$ between rigid rotation
and shear, which is depicted in
Fig.~\ref{figure_temporal_profiles}b. This provides an idea of the
deviation from cylindrical symmetry, and will be compared in the
following to the case of yield stress fluids. Note that eddies are
likely to be present in the ``rigid'' region
[\citet{Moffatt1964,Atkinson1992}]; however, we did not observe
any signature of their existence: they can thus be considered as
second-order phenomena.

\begin{figure}[htbp] \begin{center}
\includegraphics[width=9cm]{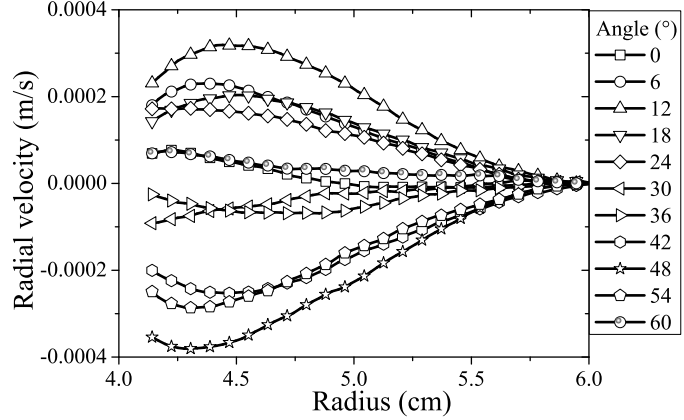}
\caption{Radial velocity profile $V_r(r,\th)$ of a Newtonian oil
sheared at 1~rpm in a six-bladed vane-in-cup geometry, for various
angular positions $\th$ between two adjacent blades (from
$\th=0$\degre to $\th=60$\degre).}\label{figure_radial_velocity}
\end{center}\end{figure}

\begin{figure}[htbp] \begin{center}
\includegraphics[width=7.25cm]{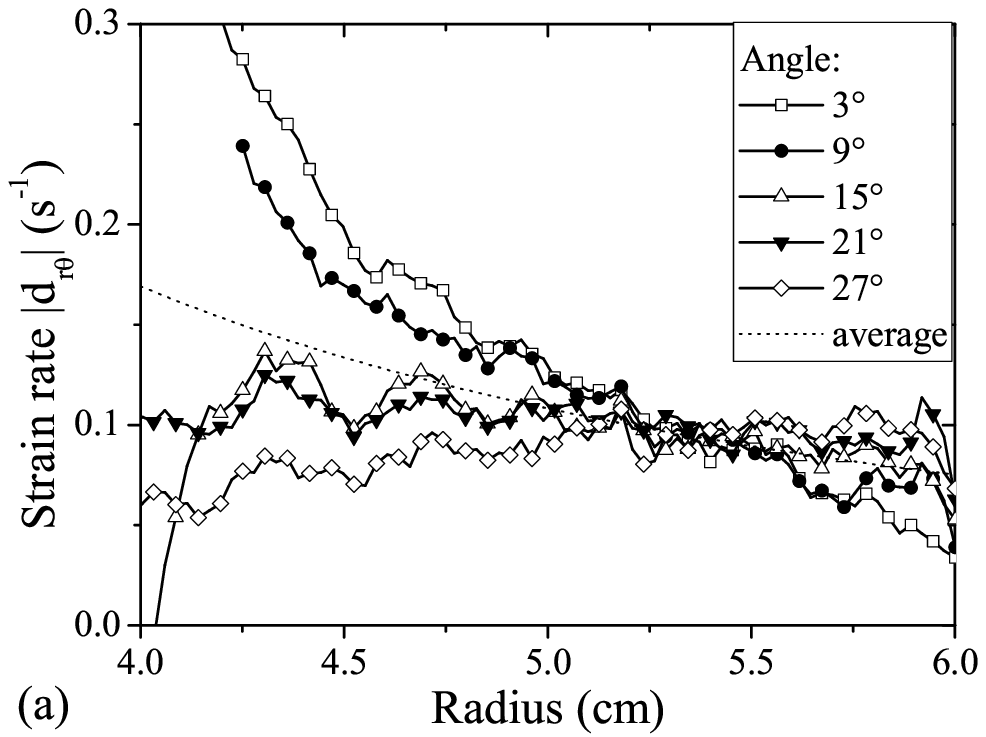}
\includegraphics[width=8.55cm]{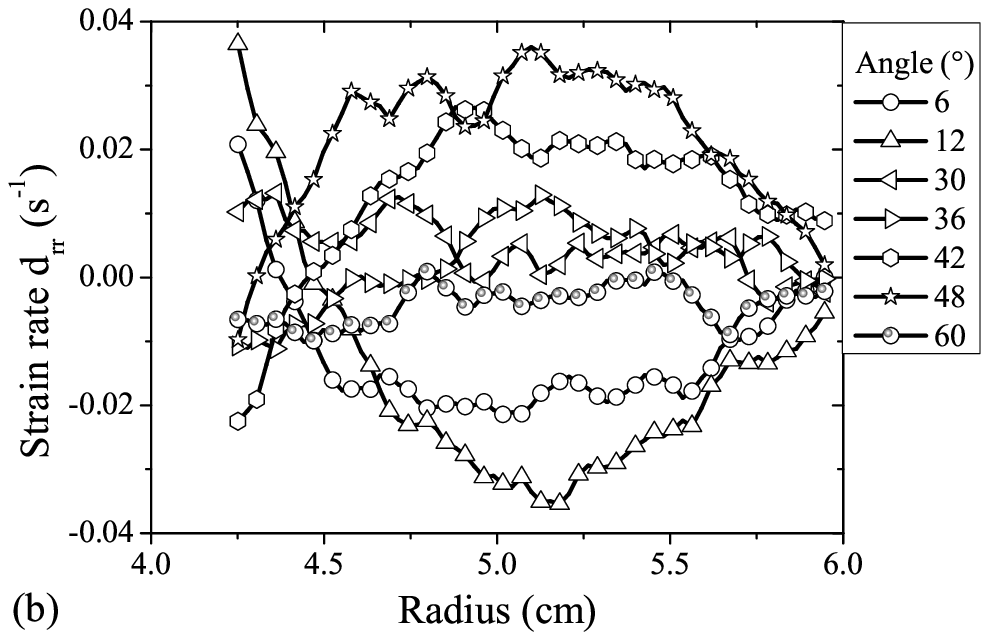}
\caption{Strain rate profiles $\drt(r,\th)$ (left) and
$\drr(r,\th)$ (right) vs. radius $r$ for various angular positions
$\th$ between two adjacent blades (from $\th=0$\degre to
$\th=60$\degre).}\label{figure_temporal_gradients}
\end{center}\end{figure}

As explained in Sec.~\ref{section_display}, from the
$V_\th(r,\th)$ measurement and from the material
incompressibility, we are able to reconstruct the radial velocity
profile $V_r(r,\th)$ (see Fig.~\ref{figure_radial_velocity}). This
also allows us to compute the strain rate components $\drt(r,\th)$
and $\drr(r,\th)=-\dtt(r,\th)$, which are plotted in
Fig.~\ref{figure_temporal_gradients}. Of course, due to the
limited number of profiles between two adjacent blades, this
method provides only a rough estimate of these quantities. In
addition to their interest for future comparison with models and
simulations, these data allow us to evaluate the contribution of
the extensional flow to dissipation; here, in a Newtonian medium,
the local power density is given by:
$p_d(r,\theta)=2\eta(\drt^2+2\drr^2)$.

In Fig.~\ref{figure_radial_velocity}, we first observe that
$V_r(r,\th)\simeq0$ for $\th=0$\degre and $\th=30$\degre; there is
thus no extensional flow in these regions of space, as seen in
Fig.~\ref{figure_temporal_gradients}. This is actually expected
from the fore-aft symmetry of the flow around these angular
positions. $V_r$ and its spatial variations (\text{i.e.} $\drr$)
are maximal at $\th\simeq15$\degre. Meanwhile, we observe that
$\drt$ is maximal near the blades: at $r\simeq R_i$ it is more
than 4 times larger at $\th=0$\degre than at $\th=30$\degre. We
then find that $\drt$ (and thus the shear stress $\trt$) decreases
more rapidly from the blades (at $\th=0$\degre) than the $1/r^2$
scaling of the Couette geometry, whereas it does not vary much
with $r$ midway between adjacent blades (it even seems to slightly
increase with $r$ as already observed in simulations by
\citet{Savarmand2007}). It is also worth nothing that at $r\simeq
R_e$, in contrast with what is observed at $r\simeq R_i$, the
shear stress value is of order two times lower at $\th=0$\degre
than at $\th=30$\degre.

From the whole set of $\drt$ and $\drr$ measurements
(Fig.~\ref{figure_temporal_gradients}), we finally find that in
regions where $\drr$ is maximal, the contribution of the
extensional flow to dissipation is of order 25\%. Over the whole
gap, we then evaluate its average contribution to dissipation to
be rather important, of order 5 to 10\%. This significant value
may be a reason why the torque that has to be exerted to enforce
flow is higher than that predicted by \citet{Atkinson1992} in an
infinite medium. The confinement effect induced by a close
boundary at a radius $R_e$ likely increases the contribution of
the extensional flow to dissipation as compared to the case of an
infinite medium (although other effects may exist, as appears from
the comparison of the data of Tab.~\ref{tab_torque_reduction}).

\subsection{Yield stress fluid}\label{section_emulsion}

In this section, we study the flows induced by the vane tool with
a yield stress fluid (a concentrated emulsion). We focus on the
behavior near the yielding transition, \textit{i.e.} on low
rotational velocities $\Omega$.

\begin{figure}[htbp] \begin{center}
\includegraphics[width=9cm]{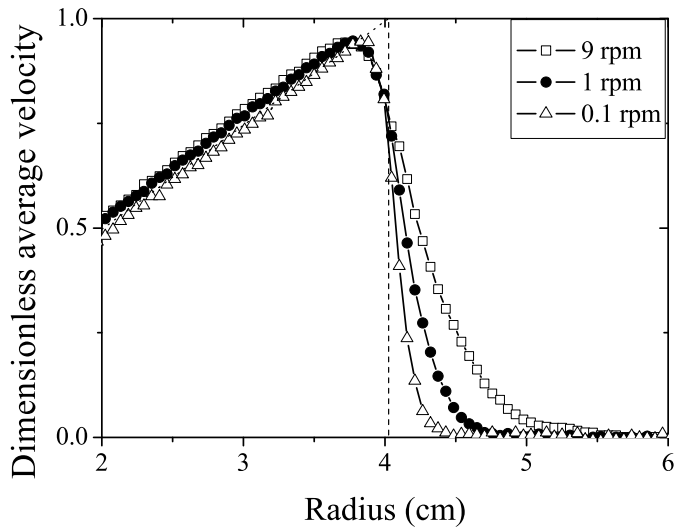}
\caption{Dimensionless $\th$-averaged velocity profile
$\barvt(r,\Omega)/\Omega R_i$ of a yield stress fluid
(concentrated emulsion) in a six-bladed vane-in-cup geometry for
$\Omega$ ranging from 0.1 to 9~rpm; the vertical dashed line shows
the radius of the vane; the dotted line is the theoretical profile
for a rigid body rotation (for
$r<R_i$).}\label{figure_average_velocity_yield}
\end{center}\end{figure}

In Fig.~\ref{figure_average_velocity_yield}, we plot the
$\th$-averaged azimuthal velocity profiles $\barvt(r)$ measured at
several $\Omega$ values ranging from 0.1 to 9~rpm, corresponding
to macroscopic shear rates varying between 0.02 and 2~s$^{-1}$. We
first observe that flow is localized: the material is sheared only
up to a radius $R_c<R_o$. $R_c$ is found to increase as $\Omega$
increases. This is a classical feature of flows of yield stress
fluids in heterogeneous stress fields. It has been observed in
Couette geometries [\citet{Coussot2005,Ovarlez2008}], where it is
attributed to the $1/r^2$ decrease of the shear stress $\trt$,
which passes below $\tau_y$ at some $R_c(\Omega)<R_o$ at low
$\Omega$. In this case, when $\Omega$ tends to 0, $R_c$ tends to
$R_i$ and the torque $T$ at the inner cylinder tends to
$\tau_y*2\pi R_i^2H$. In the vane-in-cup geometry, the same
argument holds qualitatively thanks to
Eq.~\ref{equation_tau_vane}. It implies that the flow has to stop
inside the gap at low $\Omega$. However, in contrast with the case
of the Couette geometry, as the whole stress field \textit{a
priori} depends on $\th$, this $\th$-averaged equation does not
provide the position of the limit between the sheared and the
unsheared material (which will determined at the end of this
section).

We then observe that, although this effect is less pronounced than
with a Newtonian material, the shear flow still enters the region
between the blades, even at the lowest studied $\Omega$. Close
examination of the profiles shows that the material trapped
between the blades rotates as a rigid body only up to
$R_l\simeq3.65$~cm at $\Omega=9$~rpm, $R_l\simeq3.75$~cm at
$\Omega=1$~rpm, and $R_l\simeq3.85$~cm at $\Omega=0.1$~rpm. We
recall that $R_l\simeq3.05$~cm with a Newtonian fluid in the same
geometry.

\begin{figure}[htbp] \begin{center}
\includegraphics[width=9cm]{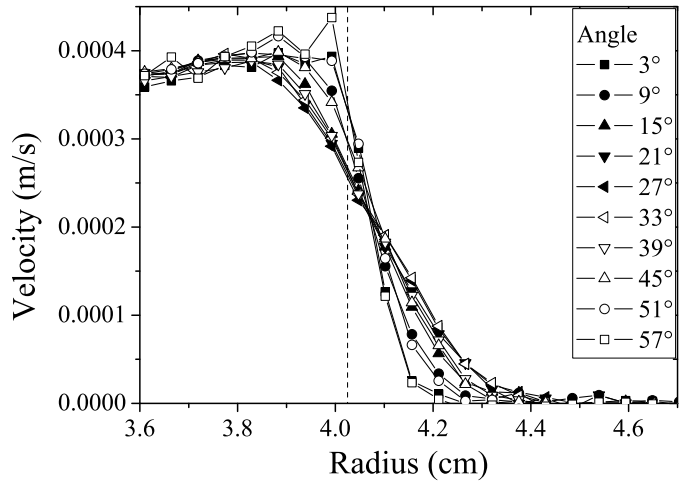}
\caption{Azimuthal velocity profile $V_\th(r,\th)$ of a yield
stress fluid (concentrated emulsion) sheared at 0.1~rpm in a
six-bladed vane-in-cup geometry, for various angular positions
$\th$ between two adjacent blades (from $\th=0$\degre to
$\th=60$\degre). The vertical dashed line shows the radius of the
vane.}\label{figure_temporal_velocity_yield}
\end{center}\end{figure}

\begin{figure}[htbp] \begin{center}
\includegraphics[width=7.9cm]{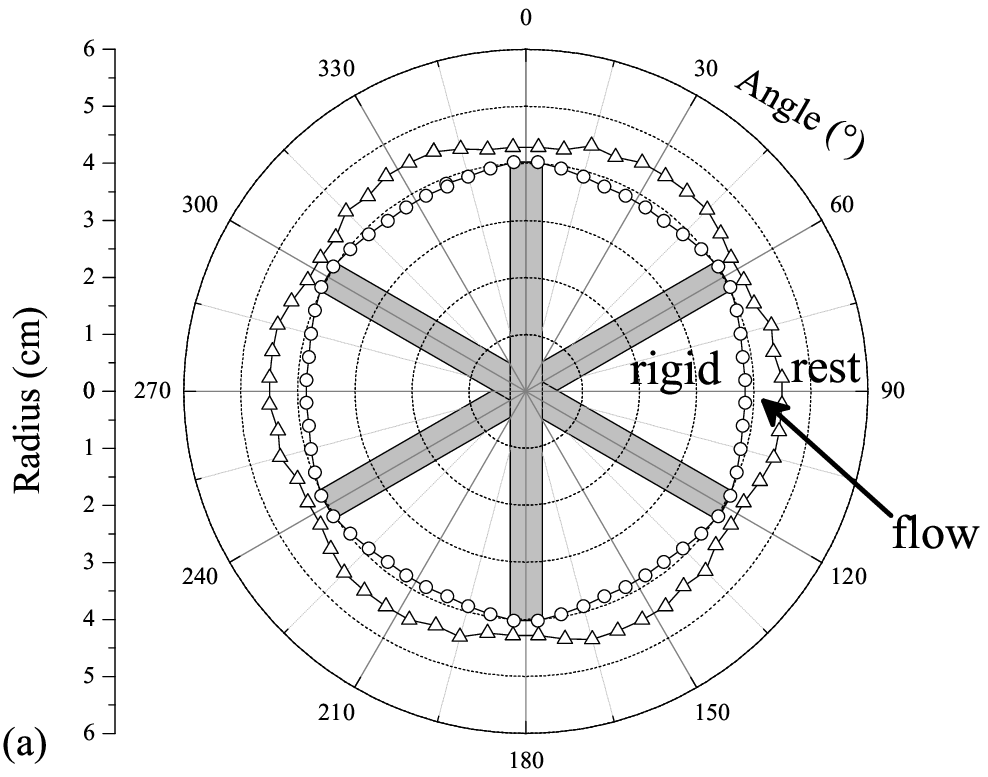}
\includegraphics[width=7.9cm]{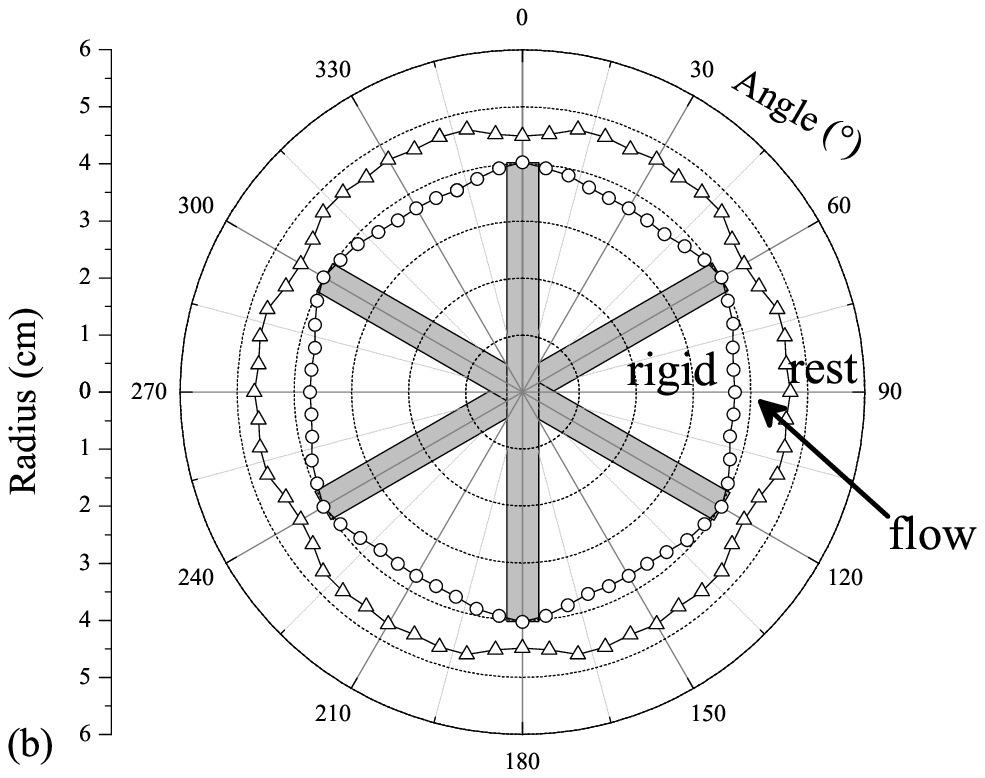}
\caption{Two-dimensional plot of the limit between rigid motion
and shear (circles) and between shear and rest (triangles) for a
yield stress fluid (concentrated emulsion) sheared in the
six-bladed vane-in-cup geometry at 0.1~rpm (left) and 1~rpm
(right). The grey rectangles correspond to the
blades.}\label{fig_rigid_limit_yield}
\end{center}\end{figure}

As in
Sec.~\ref{section_results}\ref{section_oil}\ref{section_theta_profiles},
to better characterize the flow field, we have performed
experiments in which we have measured 10 azimuthal profiles
between two adjacent blades at 0.1~rpm. In
Fig.~\ref{figure_temporal_velocity_yield}, as for a Newtonian
fluid, we observe that there is a strong $\th$-dependence of the
velocity profiles. The velocity profile that starts near from a
blade tip (at $\th=0$\degre) has a much steeper slope than the
profile measured midway between adjacent blades (at
$\th=30$\degre); again, this shows that the blade tip
neighborhoods are regions of high shear. Meanwhile the flow stops
at a radius $R_c$ which is larger at $\th=30$\degre (4.5~cm) than
at $\th=0$\degre (4.3~cm). Note also that there may be slight
fore-aft asymmetry, as sometimes observed with yield stress fluids
flows [\citet{dollet2007,Putz2008}], but we did not study this
point further. From these velocity profiles, we have reconstructed
a 2D map of the flow field (Fig.~\ref{fig_rigid_limit_yield}),
indicating both the boundary between the region of rigid body
rotation (between the blades) and the sheared region, and the
boundary between the sheared region and the outer region of fluid
at rest (\textit{i.e.} the position where the yield criterion is
satisfied).

Flow is found to occur in a layer of complex shape which is far
from being cylindrical even at this very low velocity. These
observations are in contradiction with the usually accepted
picture for yield stress fluid flows at low rates
[\cite{Barnes2001}], namely that the material contained in the
virtual cylinder delimited by the blades rotates as a rigid body,
and that it flows uniformly in a thin cylindrical layer near the
blades. Our results contrast in particular with previous numerical
works which showed that the yield surface is cylindrical at low
rates for Bingham fluids, Casson fluids, and Herschel-Bulkley
materials with $n=0.5$
[\citet{keentok1985,Yan1997,Savarmand2007}]. With apparently
similar conditions to those in some of the \citet{Yan1997}
simulations, we find an important departure from cylindrical
symmetry. This means that further investigation on the exact
conditions under which this symmetry can be recovered is still
needed. Possible difference between our work and that of
\citet{Yan1997} is that the blade thickness is zero in this last
study.

It is particularly striking and counterintuitive that $R_c$ is
largest at the angular position ($\th=30$\degre) where shear at
$R_i$ is smallest (similar observation was made by
\citet{Potanin2010}). As in
Sec.~\ref{section_results}\ref{section_oil}\ref{section_theta_profiles},
this points out the importance of the extensional flow in this
geometry, with $\th$-dependent normal stress differences which
have to be taken into account in the yield criterion, and which
thus impact the yield surface. It thus seems that the link between
the yield stress $\tau_y$ and the torque $T$ measured at yield
with a vane-in-cup geometry is still an open question, although
the classical formula probably provides a sufficiently accurate
determination of $\tau_y$ in practice.

The same 2D map as above is plotted for $\Omega=1$~rpm in
Fig.~\ref{fig_rigid_limit_yield}; the same phenomena are observed,
with enhanced departure from cylindrical symmetry, consistent with
the observation that $R_l$ decreases when $\Omega$ increases. This
result was also unexpected, as simulations find uniform flows for
shear-thinning material of index $n\leq0.5$
[\cite{Barnes1990,Savarmand2007}]; we would have expected the same
phenomenology in a Herschel-Bulkley material of index $n=0.5$ (and
thus $R_l$ to tend to $R_i$ when increasing $\Omega$). This
observation also shows that a Couette analogy can hardly be
defined for studying the flow properties of such materials in a
vane-in-cup geometry because the equivalent Couette geometry
radius $\Rieq$ would probably depend also on $\Omega$ (as recently
shown by \citet{Zhu2010}).

Let us finally note that this departure from cylindrical symmetry
has important impact on the migration of particles in a yield
stress fluid (see below).

\subsection{Concentrated suspension}\label{section_suspension}

In this section, we investigate the behavior of a concentrated
suspension of noncolloidal particles in a yield stress fluid (at a
40\% volume fraction).

A detailed study of their velocity profiles would {\it a priori}
present here limited interest: such materials present the same
nonlinear macroscopic behavior as the interstitial yield stress
fluid, and their rheological properties (yield stress,
consistency) depend moderately on the particle volume fraction
[\citet{Mahaut2008a,chateau2008}].

On the other hand, noncolloidal particles in suspensions are known
to be prone to shear-induced migration, which leads to volume
fraction heterogeneities. This phenomenon is well documented in
the case of suspensions in Newtonian fluids
[\citet{Leighton1987b,Abbott1991,Phillips1992,Corbett1995,Shapley2004,Ovarlez2006}]
but is still badly known in yield stress fluids (some studies
exist however in viscoelastic fluids
[\citet{tehrani1996,huang2000,lormand2004}]). In the model of
\citet{Leighton1987b} and \citet{Phillips1992}, migration is
related to shear-induced diffusion of the particles
[\citet{Leighton1987a,Acrivos1995}]. In a wide gap Couette
geometry, the shear stress heterogeneity is important
(Eq.~\ref{equation_tau_Couette}); the shear rate gradients then
generate a particle flux towards the low shear zones, which is
counterbalanced by a particle flux due to viscosity gradients. A
steady state, which results from competition between both fluxes,
may then be reached, and is characterized by an excess of
particles in the low shear zones of the flow geometry (near the
outer cylinder in a wide-gap Couette geometry
[\citet{Phillips1992,Corbett1995,Ovarlez2006}]). Note that there
are other models
[\citet{Nott1994,Mills1995,Morris1999,Lhuillier2009}] in which
particle fluxes counterbalance the gradients in the particle
normal stresses, and which can be used directly for non-Newtonian
media.

As the development of migration depends on the spatial variations
of shear, one may wonder how the azimuthal heterogeneities of
shear introduced by the vane tool affect migration; a related
question is that of the relevance of the Couette analogy for this
phenomenon. In the following, we thus focus on the particle volume
fraction distribution evolution when the material is sheared.

\subsubsection*{Behavior at high shear rate}

\begin{figure}[htbp] \begin{center}
\includegraphics[width=8.15cm]{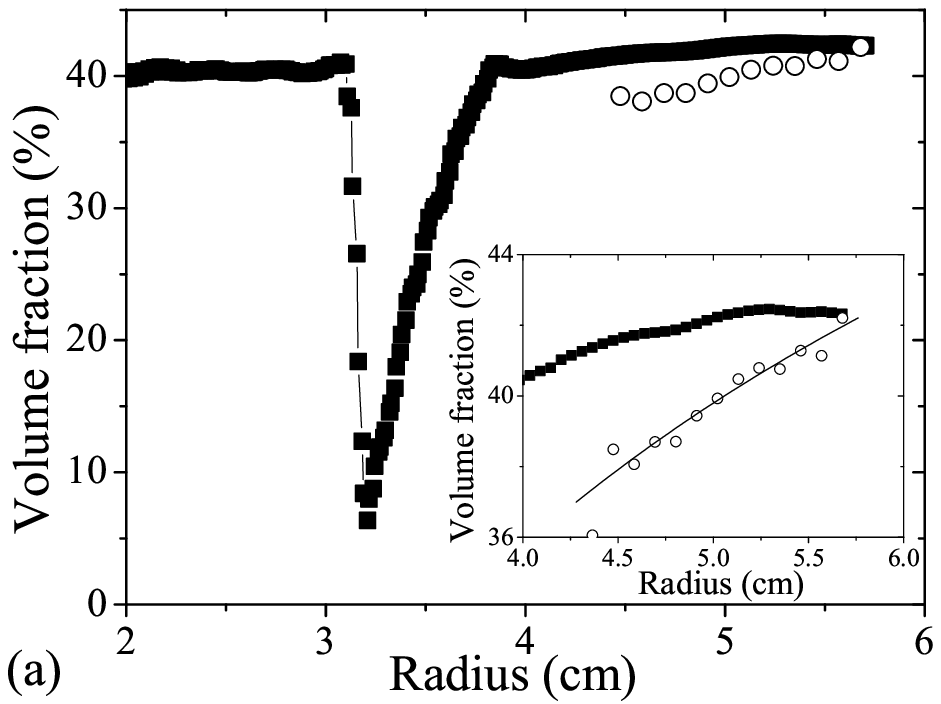}
\includegraphics[width=7.65cm]{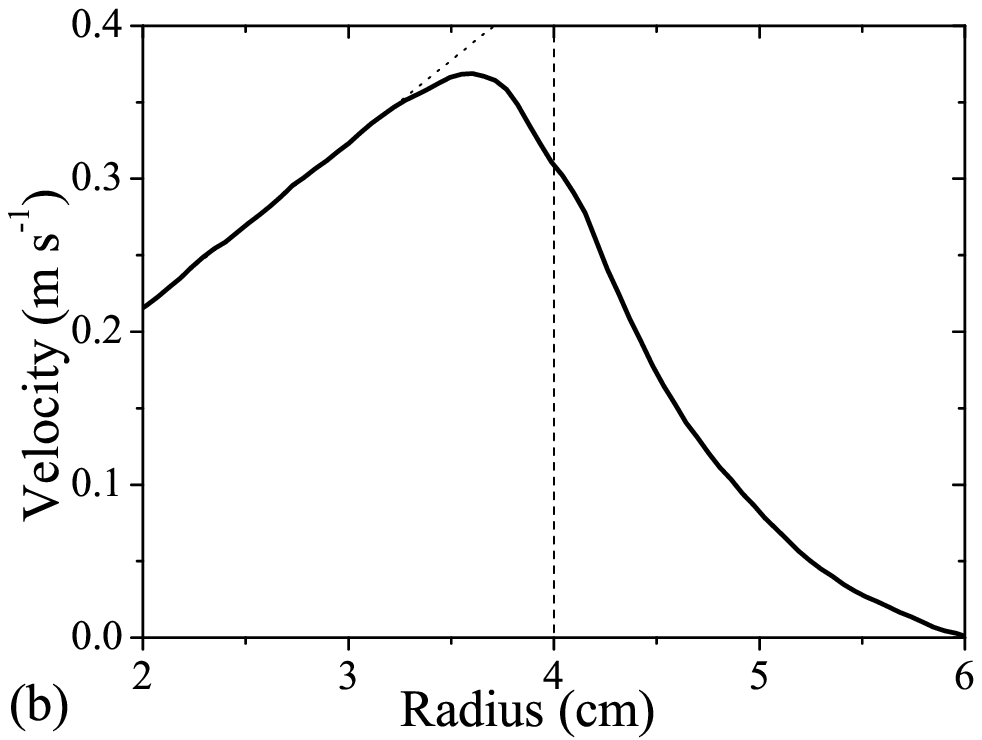}
\caption{a) Steady-state volume fraction vs. radius at
$\Omega=100$~rpm in both the Couette geometry (empty circles) and
the vane-in-cup geometry (squares). In the vane-in-cup geometry,
the volume fraction profile is determined in a 1~cm thick slice
situated exactly between two adjacent blades (see
Fig.~\ref{fig_image_vane_omega100}b). The inset is a zoom; the
line is a fit of the data measured in the Couette geometry to the
\citet{Phillips1992} model with $K_c/K_\mu=0.42$. b)
$\th$-averaged azimuthal velocity profile $\barvt(r)$; the dotted
line is the theoretical rigid motion induced by the rotation of
the vane tool; the vertical dashed line shows the radius of the
vane.} \label{fig_concentration_profile_omega100}
\end{center}\end{figure}

We first study the behavior at high shear rate, in the absence of
shear localization. We shear the suspension in both the standard
Couette geometry and the vane-in-cup geometry at a rotational
velocity $\Omega=100$~rpm. In this first set of experiments, we
only study the steady-state of migration. At $\Omega=100$~rpm,
this steady-state is reached in less than 30 min (which
corresponds to a macroscopic strain of order 50000, consistently
with strainscale evaluations from data of the literature
[\citet{Ovarlez2006}]). In
Fig.~\ref{fig_concentration_profile_omega100} we plot the steady
state volume fraction profiles observed after shearing the
suspension at $\Omega=100$~rpm during 1h.

As expected, we first observe that the material is strongly
heterogeneous in the Couette geometry: the volume fraction varies
between 37\% near the inner cylinder and 43\% near the outer
cylinder (note that the NMR technique we use do not allow
quantitative measurements near the walls). This heterogeneity is
quantitatively similar to that observed in Couette flows of
Newtonian suspensions at a same 40\% particle volume fraction
[\citet{Corbett1995}]; the profiles are actually well fitted to
the \citet{Phillips1992} model (see Eq.~16 of \citet{Ovarlez2006})
with a dimensionless diffusion constant $K_c/K_\mu=0.42$ which is
close to that found by \citet{Corbett1995} ($K_c/K_\mu=0.36$),
although this model is not expected to hold in non-Newtonian
suspensions.

\begin{figure}[htbp] \begin{center}
\includegraphics[width=12cm]{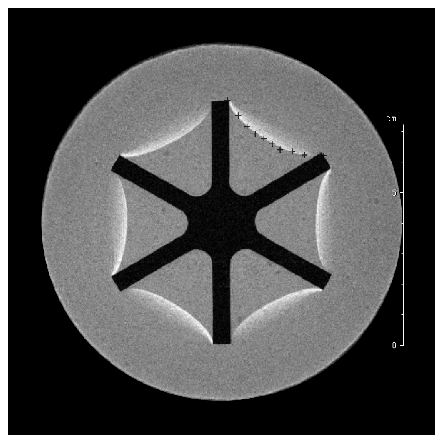}\\
(a)\\ \ \\
\includegraphics[width=5.9cm]{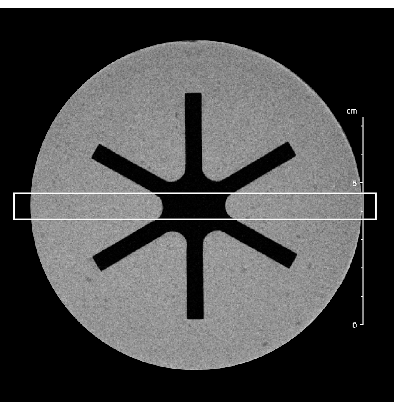}
\includegraphics[width=5.91cm]{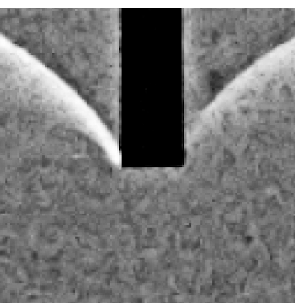}\\
(b)\hfil(c) \caption{2D magnetic resonance image of a suspension
of particles in a yield stress fluid in a vane-in-cup geometry:
(a) after a 1h shear at $\Omega=100$~rpm (corresponding to a
macroscopic strain of order 75000), and (b) before any shear. The
crosses in Fig.~\ref{fig_image_vane_omega100}a correspond to the
limit between rigid motion and shear for the Newtonian oil of
Fig.~\ref{figure_temporal_profiles}b. The white rectangle in
Fig.~\ref{fig_image_vane_omega100}b shows the slice in which the
volume fraction profiles of
Figs.~\ref{fig_concentration_profile_omega100}a and
\ref{fig_concentration_profile_omega1} are measured. (c) is a zoom
of image (a) near the edges of a blade. The images are taken in
the horizontal plane of the geometry, at middle height of the vane
tool, and correspond to vertical averages over 2~cm. The vane tool
rotates counterclockwise.}\label{fig_image_vane_omega100}
\end{center}\end{figure}

In the vane-in-cup geometry, the volume fraction profile shows
very different features; note that the profile is measured in a
1~cm thick (in the azimuthal direction) slice situated exactly
between two adjacent blades (see
Fig.~\ref{fig_image_vane_omega100}). In
Fig.~\ref{fig_concentration_profile_omega100}, we first observe
that there is a strong particle depletion in a wide zone between
the blades. A homogeneous volume fraction of 40\% is observed for
radii inferior to 3.1~cm. At a radius $R_l=3.1$~cm, there is a
strong drop in the volume fraction down to 5\% within 1~mm
(corresponding to 4 particle diameters). Close inspection of the
velocity profile Fig.~\ref{fig_concentration_profile_omega100}
shows that this radius $R_l$ corresponds to the transition between
rigid motion and shear between the blades. The volume fraction
then increases basically linearly with the radius up to a 40.5\%
volume fraction at a radius $r=3.85$~cm which is close to the vane
radius. The volume fraction then increases only slightly (between
40.5\% and 42.5\%) in the gap of the geometry: the heterogeneity
is here much less important than in a standard Couette geometry.

To get further insight into the new strong depletion phenomenon we
have evidenced, we have performed 2D magnetic resonance images of
the material. Such images provide a qualitative view of the
spatial variations of the particle volume fraction as only the
liquid phase is imaged. Images are coded in grey scales; a
brighter zone contains less particles. In
Fig.~\ref{fig_image_vane_omega100}b, we first see an image of the
homogeneous material. Before any shear, as expected, the light
intensity is homogeneous in the sample (intensity variations
correspond to noise). After a 1h shear at $\Omega=100$~rpm, we
observe very bright and thin curves on the image: they correspond
to zones where the volume fraction suddenly drops down to a value
close to zero. These curves are not circles. More precisely,
between two adjacent blades, a depleted zone goes from the edge of
one blade (at $\th=0$\degre, $r=4.02$~cm) to the edge of another
blade (at $\th=60$\degre, $r=4.02$~cm), and describes a concave
$r(\theta)$ curve whose minimum is $r=3.1$~cm at $\th=30$\degre.
Note that as the volume fraction profile is averaged over a slice
which is 1~cm thick in the azimuthal direction (see
Fig.~\ref{fig_image_vane_omega100}b), the fact that we measure a
minimum of 5\% at $r=3.1$~cm in the slice probably means that the
volume fraction minimum is actually equal to zero in the depletion
zone.

As pointed out above, this curve also likely marks the transition
between the unsheared material (which rotates as a rigid body) and
the sheared material. Note in particular the similarity with
Fig.~\ref{figure_temporal_profiles}b, the data of which are
reported in Fig.~\ref{fig_image_vane_omega100}a for illustration.
A first interpretation of the phenomenon would then simply be that
migration is caused by shear and naturally stops at this
transition zone. Indeed, as shear is maximum near the blades,
particles tends to migrate out of this zone; moreover, there is no
source of particle flux from the unsheared zone between the blades
to balance the migration towards the outer cylinder. However, this
does not explain why the volume fraction drops down to zero:
heterogeneities observed at steady-state in the literature are
usually moderate and do not lead to zones free of particles. A
better understanding of the phenomenon can be gained by zooming on
the previous image (Fig.\ref{fig_image_vane_omega100}c). We now
see that while the depletion phenomenon seems symmetric around
both sides of the blades at a macroscopic scale, it is clearly
asymmetric at a local scale near the blades and depends on the
direction of rotation: depletion is more pronounced at the back of
the blade (note that the vane tool rotates counterclockwise). This
would mean that the noncolloidal particle trajectories are
asymmetric around the blade: a particle that is found at a radius
$r\simeq 4.02$~cm just before the blade is necessarily found at a
radius slightly higher than 4.02~cm after the blade as there are
no particles at $r=4.02$~cm. This feature is reminiscent of the
fore-aft asymmetry that is observed in the bulk of noncolloidal
suspensions [\citet{parsi1987}] and that leads to their
non-Newtonian properties [\citet{Brady97}]. It thus seems that, in
addition to the shear-induced migration mechanism intrinsic to
suspensions, the vane tool induces a specific migration mechanism
which has its origin in the direct interactions between the
particles and the blades; this effects leads to the full depletion
that is observed at the transition between the sheared and the
unsheared material. Such direct effect of a flow geometry on
migration has also been observed in microchannel flows of
colloidal suspensions [\citet{Wyss2006}], and also led to full
particle depletion. See also \citet{Jossic2004}. The kinetics of
the phenomenon will be briefly discussed below.

The rest of the volume fraction profile results from a complex
interplay between shear-induced migration and the fore-aft
asymmetry around the blades; this leads to the rapid increase of
the volume fraction between 3.1~cm and 4.02~cm. After 4.02~cm the
flow lines do not meet the blade edges, and the phenomenon
evidenced above should have basically no effect on the
heterogeneity that develops in the gap of the geometry. On the
other hand, the mean volume fraction should be slightly higher due
to mass conservation; it is indeed observed to be equal to around
42\%. Nevertheless, as the mean radial shear rate heterogeneity is
basically similar to that observed in a standard Couette geometry
(see previous sections), we would {\it a priori} expect the
heterogeneity to be somehow similar. However, we observe that the
volume fraction profile is only slightly heterogeneous: there is
less than 5\% variation of the volume fraction in the gap, to be
compared to the 15\% variation observed in the Couette geometry.
Clearly, this means that the Couette analogy is irrelevant as
regards this phenomenon, and that the details of shear matter.
Here, the extensional flow that adds to shear may be at the origin
of this diminution of migration. A more detailed analysis is out
of the scope of this paper.

\subsubsection*{Behavior at low shear rate}

\begin{figure}[htbp] \begin{center}
\includegraphics[width=9cm]{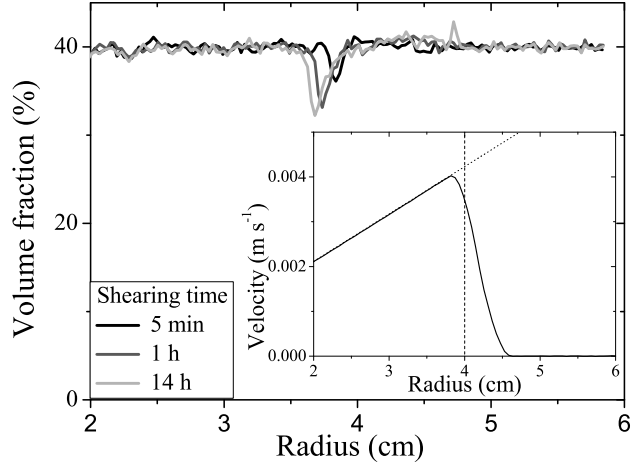}
\caption{Volume fraction vs. radius at $\Omega=1$~rpm measured in
the vane-in-cup geometry after different times of shear: 5min, 1h,
14h. The material is homogeneous at the beginning of shear. The
inset presents the $\th$-averaged azimuthal velocity profile
$\barvt(r)$ measured in the first stages of shear; the dotted line
is the theoretical rigid motion induced by the rotation of the
vane tool; the vertical dashed line shows the radius of the vane.}
\label{fig_concentration_profile_omega1}
\end{center}\end{figure}

Let us now study the behavior at low shear rate. Low shear rates
are typically imposed with the aim of measuring the yield stress
of such materials. Starting from a homogeneous suspension at rest,
we apply a rotational velocity $\Omega=1$~rpm (without any
preshear), and we measure the evolution of the particle volume
fraction in time. The corresponding volume fraction profiles are
depicted in Fig.~\ref{fig_concentration_profile_omega1}.

In Fig.~\ref{fig_concentration_profile_omega1}, we observe that,
although shear is much less important than in the previous
experiments, particle depletion also appears between the blades.
Comparison of the velocity profile and the volume fraction profile
shows that depletion also appears between the blades at the
transition zone between the sheared and the unsheared materials.
This phenomenon appears with a very fast kinetics: the lower
volume fraction value in the measurement zone is 36\% after only a
5 minute shear (corresponding to a macroscopic strain of order
50). Afterwards, it continues evolving slowly: the minimum
observed volume fraction is of order 33\% after a 1h shear and of
order 32\% after a 14h shear (corresponding to a 10000 macroscopic
strain). Note that the radial position of the minimum value of the
volume fraction slightly decreases in time; it likely corresponds
to progressive erosion of the material between the blades (we did
not measure the velocity profiles to check this hypothesis).

We also note that migration is negligible in the rest of the
sheared material as expected from the theory of migration briefly
described above (a larger strain would be needed to observe
significant migration). Nevertheless, we note some particle
accumulation (with a volume fraction value of 43\%) at
$R_c$=4.7~cm after a very long time. This corresponds to the yield
surface as flow is localized at low velocity (see velocity profile
Fig.~\ref{fig_concentration_profile_omega1}). Migration profiles
usually result from an equilibrium between various sources of
fluxes. On the other hand, the unsheared material does not produce
any particle flux while it receives particles from the sheared
region. This particle accumulation is thus the signature that the
migration phenomenon is indeed active, although not observable on
the profile measured in the sheared zone. It is probable that this
accumulation process would stop only (after a very long time) when
there are no more particles in the sheared region.

\begin{figure}[htbp] \begin{center}
\includegraphics[width=7.9cm]{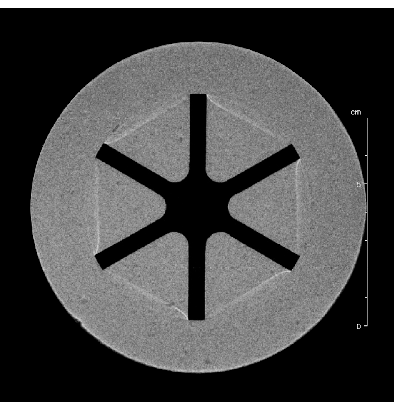}
\includegraphics[trim=0cm 0.4cm 0cm
0.045cm,clip,width=7.9cm]{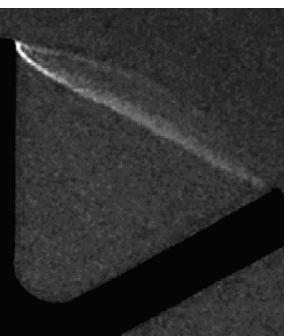}\\(a)\hfill(b) \caption{(a) 2D
magnetic resonance image of a suspension of particles in a yield
stress fluid in the horizontal plane of a vane-in-cup geometry
after a 14h shear at $\Omega=1$~rpm. (b) is a zoom of image (a)
between two adjacent blades. The image is taken in the horizontal
plane of the geometry, at middle height of the vane tool, and
corresponds to a vertical average over 2~cm. The vane tool rotates
counterclockwise.}\label{fig_image_vane_omega1}
\end{center}\end{figure}

As above, 2D magnetic resonance images of the material provide an
insight in the phenomenon. In Fig.~\ref{fig_image_vane_omega1}, we
observe again that particle depletion is enhanced at the rear of
the blades; this confirms that this phenomenon is likely due to
direct interactions between the blades and the particles, leading
to the asymmetry of the particles trajectory around the blades.
This \textit{a priori} occurs with any particle whose trajectory
is close to the blades, explaining why particle depletion appears
so rapidly. There is probably no way to avoid it. Note that the
images are here much brighter very close to the blades than midway
between two adjacent blades; this would mean that the particle
volume fraction is probably
close to $0$ near the blades, although we observe volume fraction of order 32\% between two blades.\\

Finally, let us note that the bright line provides a good idea of
the boundary between the sheared material and the material that
moves as a rigid body. We see as in
Sec.~\ref{section_results}\ref{section_emulsion} that this is far
from being cylindrical even at this low velocity.

\subsubsection*{Consequences: slip with a vane tool}

We finally present some consequences of this phenomenon. From the
above observations, our conclusion is that depletion sets up
quickly and is probably unavoidable. Then two situations have to
be distinguished. If linear viscoelastic properties of a
suspension of large particles are measured at rest on the
homogeneous material in its solid regime, without any preshear,
then these measurements pose no other problem than that of the
relevant Couette analogy to be used (see
Sec.~\ref{section_results}\ref{section_oil}). If a yield stress
measurement is performed at low imposed rotational velocity,
starting from the homogeneous material at rest, then this
measurement is likely valid as long as only the peak value or the
plateau value at low strain (of order 1) is recorded. On the other
hand, any subsequent analysis of the material behavior will {\it a
priori} be misleading: irreversible changes have occurred and the
material cannot be studied anymore. More generally, any
measurement performed after a preshear will be incorrect and any
flow curve measurement will lead to wrong evaluation of the
material properties. In these last cases, the consequence of the
new particle depletion phenomenon we have evidenced is a kind of
wall slip near the blades, whereas there are no walls. Here the
``slip layer'' is made of the (pure) interstitial yield stress
fluid in a zone close to the blades, as would be observed near a
smooth inner cylinder. This contrasts with the common belief that
the vane tool prevents slippage.

\begin{figure}[htbp] \begin{center}
\includegraphics[width=8cm]{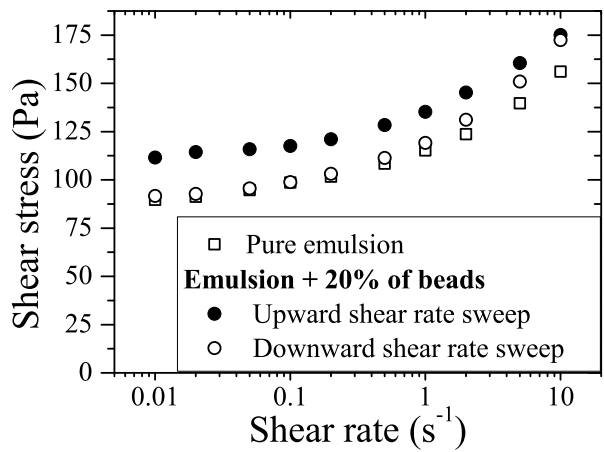}
\caption{Shear stress vs. shear rate in a vane-in-cup geometry for
upward/downward shear rate sweeps in a pure concentrated emulsion
(open squares) and for the same emulsion filled with 20\% of
140~$\mu$m PS beads (filled/open circles). Figure from
\citet{Mahaut2008a}.}\label{fig_sweep_suspension}
\end{center}\end{figure}

In order to illustrate this feature, we present some results of
\citet{Mahaut2008a}: Mahaut \textit{et al.} performed classical
upward/downward shear rate sweeps with a six-bladed vane-in-cup
geometry ($R_i=1.25$~cm, $R_o=1.8$~cm, $H=$4.5~cm, blade
thickness=0.8~mm) in a pure concentrated emulsion, and in the same
emulsion filled with 20\% of 140~$\mu$m PS beads. In these
experiments, constant macroscopic shear rates steps increasing
from 0.01 to 10~s$^{-1}$ and then decreasing from 10 to
0.01~s$^{-1}$ were applied during 30s, and the stationary shear
stress was measured for each shear rate value. The results are
shown in Fig.~\ref{fig_sweep_suspension}.

While the same curve is observed for the upward/downward shear
rate sweeps in the case of the pure emulsion (as expected for a
simple non-thixotropic yield stress fluid), the shear stress
during the upward shear rate sweep differs from the shear stress
during the downward shear rate sweep in the case of the
suspension. Moreover, any measurement performed on the suspension
after this experiment gives a static yield stress equal to the
dynamic yield stress observed during the downward sweep. This
means that there has been some irreversible change. This
irreversible change is actually the particle depletion near the
blades we have observed in this paper. As the flow of the
suspension is localized near the inner tool at low shear rate, it
means that after the first upward sweep that has induced the
particle depletion, during the downward shear rate sweep only the
pure emulsion created by migration near the blades remains in the
sheared layer at sufficiently low rotational velocity. This
explains why the same apparent value of the yield stress is found
in the suspension during the downward sweep as in the pure
emulsion with this experiment. On the other hand, the yield stress
at the beginning of the very first upward sweep is that of the
suspension as migration has not occurred yet.

The conclusion is that the vane tool is probably not suitable to
the study of flows of suspensions of large particles.

\section{Conclusion}
As a conclusion, let us summarize our main findings:
\begin{itemize}
\item In the case of Newtonian fluid flows, our measurements
support the Couette equivalence approach: the $\th$-averaged
strain rate component $\drt$ decreases as the inverse squared
radius in the gap. Interestingly, the velocity profiles allow
determining the Couette equivalent radius without end-effect
correction and independently of the viscosity of the material. The
torque exerted by the vane in our display is found to be higher
(by 8\%) than the theoretical prediction of \citet{Atkinson1992}
for a vane embedded in an infinite medium, and is thus much closer
to the torque exerted by a Couette geometry of same radius as the
vane than expected from the literature. A key observation may be
that there is a significant flow between the blades which adds an
important extensional component to shear, thus increasing
dissipation. From a short review of the literature, it clearly
appears that numerical investigations are still needed in the case
of finite geometries. Variational approaches are also promising,
although they do not yet provide tight bounds. \item In the case
of yield stress fluid flows, we find that the thin layer of
material which flows around the vane tool at low velocity is not
cylindrical, in contrast with what is usually supposed in the
literature from simulation results. Consequently, a non negligible
extensional component of shear has probably to be taken into
account in the analysis. At this stage, there are too few
experimental and simulation data to understand the origin of this
discrepancy. It thus seems that progress still has to be made, in
particular through simulations, which allow a wide range of
parameters to be studied. This may help understanding how the
torque is linked to the yield stress of a material at low
velocity, depending in particular on the geometry. \item An
important and surprising result is the observation of particle
depletion near the blades when the yield stress fluid contains
noncolloidal particles. This phenomenon is thus likely to occur
when studying polydisperse pastes like coal slurries, mortars and
fresh concrete. It has to be noted that the phenomenon is very
rapid, irreversible, and thus probably unavoidable when studying
flows of suspensions of large particles. It results in the
creation of a pure interstitial yield stress fluid layer and thus
in a kind of wall slip near the blades. It contrasts with the
classical assumption that is made in the field of concentrated
suspension rheology where the vane tool is mainly used to avoid
this phenomenon.
\end{itemize}

\noindent Consequently, we would say that, in the case of pasty
materials with large particles, if accurate measurements are
needed, the vane tool may finally be suitable only for the study
of the solid (elastic) properties of materials and for the static
yield stress measurements; as the yield stress measurement may
induce irreversible particle depletion near the blades, any new
measurement then requires a new sample preparation. Furthermore,
the vane can be used as a very accurate tool without any
hypothesis nor any calibration to measure the relative increase of
the elastic modulus of materials as a function of their
composition [\citet{alderman1991,Mahaut2008a}]. In order to study
accurately the flows of pasty materials with large particles, our
results suggest that a coaxial cylinders geometry with properly
roughened surfaces is preferable when possible. If the use of a
vane tool cannot be avoided, one should keep in mind our
observations in order to carefully interpret any result.

\end{document}